\documentstyle[prc,epsf,aps]{revtex}
\begin{document}
\title{Polarization degrees of freedom in 
photoinduced two-nucleon knockout from finite nuclei}
\author{Jan Ryckebusch}
\address{Department of Subatomic and Radiation Physics \protect\\
University of Gent, Proeftuinstraat 86, B-9000 Gent, Belgium
\protect \\ and \protect \\ 
Institute for Nuclear Theory, University of Washington,  
Box 351550, \protect \\ Seattle, WA 98195}
\author{Dimitri Debruyne and Wim Van Nespen }
\address{Department of Subatomic and Radiation Physics \protect\\
University of Gent, Proeftuinstraat 86, B-9000 Gent, Belgium}
\date{\today}
\maketitle
\begin{abstract}
The polarization degrees of freedom in photoinduced two-nucleon
knockout from finite nuclei are studied.  It is pointed out that they
open good perspectives to study the dynamics of dinucleons in the
medium in detail.  The ($\gamma,pp$) and ($\gamma,pn$) angular cross
sections, photon asymmetries and outgoing
nucleon polarizations
are calculated for the target nuclei $^{16}$O and
$^{12}$C and photonenergies ranging from 100 up to 500~MeV.  It is
investigated to which degree the two-nucleon emission reaction is
dominated by photoabsorption on $^3S_1(T=0)$ proton-neutron and $^1S_0(T=1)$
proton-proton pairs in the nuclear medium.  The calculations
demonstrate  that dominance of $S$ wave photoabsorption 
in the ($\gamma,pn$) channel does not
necessarily imply that the reaction mechanism is similar to what is
observed in deuteron photodisintegration.

\end{abstract}
\section{Introduction}
In the late fifties it was pointed out by K. Gottfried that after
making a few approximations, the cross section for photoinduced two-nucleon ($2N$)
knockout $\sigma (\gamma,N_a N_b)$ can be related to
quantities that are sensitive to the relative and center-of-mass
motion of nucleon pairs in the nuclear system \cite{gottfried}.  In
this pioneering work, it was predicted that $2N$ knockout cross
sections are proportional to the so-called pair function $F(P)$, which
is related to the probability of finding a nucleon pair with c.m. momentum
$P$ in a finite nuclear system.  The c.m. 
momentum $P$ can be determined in
terms of the momenta of the two escaping nucleons ($\vec{k}_a$ and
$\vec{k}_b$) and the momentum transferred by the photon ($\vec{q}_{\gamma}$)~:
\begin{equation}
\vec{P} = \vec{k} _a + \vec{k} _b - \vec{q} _{\gamma}\; .
\end{equation}  
It should be noted that as soon as one goes beyond the approximations
that are at the basis of the Gottfried approach, the differential
cross sections can no longer be formally factorized in terms of the
pair function and factors that depend on the relative motion of the
active pair.  This is for example the case when considering more
realistic wave functions than plane waves for the outgoing nucleons.
Nevertheless, the dominant role of the pair function $F(P)$ in the
A$(\gamma,pn)$ and A$(\gamma,pp)$ processes has been experimentally
confirmed \cite{early,dancer,harty,peter}.  This observation raises
confidence in that the $(\gamma,N_a N_b)$ reaction mechanism can be
sufficiently kept under control to gain also empirical information
about the relative motion of nucleon pairs in the medium. It is worth
noting that also the measured $^{12}$C$(e,e'pp)$
\cite{guenther} and $^{16}$O$(e,e'pp)$ \cite{gerco} cross sections
have recently been shown to scale in terms of the pair function
$F(P)$.

Pioneering two-nucleon knockout studies with moderate energy
resolution were performed at Tokyo and Bonn
\cite{tokyo,mainz}.  In recent years, the PIP-TOF collaboration has
collected a vast amount of 2N knockout data at the tagged photon
facility of the 800~MeV Mainz electron accelerator.  In comparison
with earlier work, these measurements are characterized by an improved
energy resolution and cover a wide photon energy region ($100 \leq E_
{\gamma} \leq 800$~MeV).  Measurements with polarized photon beams on
$^4$He and $^{12}$C are in the process of being analyzed.  Other
experiments with polarized photon beams were performed for the target
nuclei $^{3}$He \cite{tede} and $^{16}$O \cite{legs} at the LEGS
facility located at the Brookhaven National Laboratory.  In these
experiments, the photon energy ranged from 220 up to 305 MeV.  The
$^3$He($\vec{\gamma},pn)p$ asymmetries turned out to be remarkably
similar to those obtained from d($\vec{\gamma}$,$p$)$n$, provided that
one selects that part in the proton-neutron
phase space $d \Omega _p d \Omega _n d
E_p d E _n$ for which the residual proton
$p$ can almost be guaranteed to be at rest.  With this restriction, only
two nucleons are involved in the reaction process and the residual proton
is a spectator.  Recently, it was pointed out by Sandorfi and
Leidemann \cite{sandorfi} that the mere assumption of proton pairs
predominantly moving in a relative $^1S_0$ state has major implications
for the linear-polarization $^3$He($\vec{\gamma},pp)n$ asymmetries.
Along the same lines, Wilhelm, Niskanen and Arenh\"{o}vel
\cite{wilh} pointed out that relatively simple forms for the $(\gamma,pp)$
cross sections and polarization observables can be obtained when
assuming photoabsorption on $^1S_0$ diprotons and dominance of any two 
of the three multipoles $(E1,E2,M2)$ in the transition matrix elements.   


Turning to finite nuclei, a number of
models to deal with two-nucleon knockout have been developed over the
last couple of years.  The
models fall into different categories.  The one developed by the
Valencia group \cite{carrasco} aims at describing all the different
aspects of the photon-nucleus coupling and pion final-state
interactions in a unified diagrammatic approach.  This includes all
the (multi-)pion and (multi-)nucleon production channels.  A very
large amount of contributing mechanisms are included in the
description of the pion and nucleon photoproduction channels.  All
this at the expense of the nuclear structure aspects of the reaction
which are dealt with in a nuclear-matter approach.  The local-density
approximation (LDA) is applied to obtain 
results for finite nuclei. The
Valencia calculations have been carefully compared to the
$(\gamma,NN)$ data collected in recent years
\cite{cross,salad,zeitschrift}.  Generally a fair description of the
missing energy spectra is obtained, an exception made for the
$(\gamma,pp)$ channel for which the calculations tend to overestimate
the data \cite{cross,salad}.  
A major conclusion from these comparisons is that for the
photoabsorption mechanism to be two-body in nature one has to select
these processes for which the $A-2$ fragment is created at low missing
energies \cite{harty}.  At higher missing energies the two-nucleon
knockout strength is predicted to be essentially due to initial
$(\gamma,\pi)$ production and subsequent pion final-state
interactions. This implies that the higher missing energy range is
excluded when studying the properties of dinucleons in the medium.
Indeed, in these studies it is essential to guarantee that only two
nucleons participate in the reaction process.

The models developed in Pavia \cite{pavia} and Gent \cite{gent} are
less ambitious as far as the number of included reaction mechanisms
is concerned.  They  concentrating on the two-body
photoabsorption mechanisms and do not include 
the pion production channels.   In the light of the previous discussion,
these models are therefore restricted to the low missing energy part
of the two-nucleon knockout spectra.  In comparison with the Valencia
approach, the Gent and Pavia models put more emphasis on a proper
description of the nuclear structure aspects of the 2N emission
reaction process.  In both models, the shell-model framework is the
starting point to account for the nuclear structure aspects.  In the
light of 2N emission processes being a possible probe to study
ground-state correlations, the mean-field wave functions are corrected
for (central) Jastrow correlation effects.  

Here we report on calculations that aim at investigating the dynamics 
of dinucleons in the medium with the aid of photoinduced two-nucleon
knockout. More in particular, we concentrate on the additional degrees
of freedom created by using (linearly) polarized photon beams and
recoil nucleon polarimetry.  The organization of this paper is as
follows. In Sect.~\ref{sec:ang} the definitions and conventions
 for the $(\gamma,NN)$ cross sections and polarization observables are given.
Sect.~\ref{sec:model} is devoted to a description of the different
assumptions which are at the basis of the model calculations
presented here.   A factorized (Subsect.~\ref{sec:fac}) and
an unfactorized (Subsect.~\ref{sec:unfac}) model are sketched.  Rather
than elaborating on the technical details, we make an attempt to
describe the basic ideas and assumptions behind the models.  
Section \ref{sec:results} contains the results of the numerical
calculations.  We concentrate on proton-proton and proton-neutron
knockout from the target nuclei $^{16}$O and $^{12}$C in several
kinematical conditions and consider photon energies from 100 up to 500~MeV.   
Finally, Sect.~\ref{sec:conclus} contains the conclusions.

\section{Differential Cross sections and polarization observables}
\label{sec:ang}
Using standard techniques the  
four-fold differential cross section for the photoinduced
two-nucleon knockout process $A+ \gamma \longrightarrow A-2 + 
N_a +N_b$ in the
LAB frame ($\vec{p}_A = \vec{o}$) reads 
\begin{equation}
\frac {d^4 \sigma}
{d \Omega _a d \Omega _b dE_a dE_b}
= \frac {1} {(2\pi)^5 E_{\gamma}} k_a k_b E_b E_a 
\delta (E_{A-2} + E_a + E_b - E_A - E_{\gamma})
\overline {\sum _{fi}} \mid m_F^{fi} (\lambda) \mid^2 \;,
\label{eq:cross}
\end{equation}
where $\overline {\sum} _{fi}$ denotes the appropriate averaging over
the inital states and sum over the final states.  The variables $k_a$
($k_b$) and $E_a$ ($E_b$) refer to the three-momentum and total energy
of the escaping nucleons.  Further, $\lambda$ denotes the photon circular
polarization.
The $m_F^{fi}(\lambda)$ is the transition matrix element for the reaction
under study and is defined as 
\begin{equation}
m_F^{fi} (\lambda = \pm 1)
 =  \left< \Psi_{f}^{(A-2)}(E_x,J_R M_R);
{\vec k}_a m_{s_{a}};{\vec k}_b m_{s_{b}}
\mid J_{\lambda} ({\vec q} _ {\gamma}) \mid \Psi _0 \right> \; ,
\label{mfia}
\end{equation}
where  $\mid {\vec k}_a m_{s_{a}} >$ and $\mid {\vec k}_b m_{s_{b}} >$
are the distorted waves of the outgoing nucleon $N_a$ and $N_b$
respectively.   The 
$\mid \Psi_{f}^{(A-2)}(E_x,J_R M_R) > $
determines the wave function of the residual $A-2$ system.  Its
excitation energy $E_x$ is expressed relative to the ground-state
energy of the bound $A-2$ nucleus.  In all forthcoming derivations we assume
that the residual nucleus is created in a specific state
characterized by an excitation energy $E_x$ and angular momentum
$J_R$.  This does not imply, however, that our considerations are
restricted to the discrete part of the $A-2$ spectrum.  Indeed, when
the situation occurs that the $A-2$ nucleus is created in the continuous
part of its spectrum a summation over the momenta $J_R$ has to be carried 
out.

In all further considerations, 
the z axis is chosen along the direction of the
incoming photon momentum.  The $xz$ plane is defined by $\vec{q}_{\gamma}$ and
$\vec{k}_a$.    The
y axis is then along the direction of the vector $ \vec{q} _{\gamma} \times
\vec{k} _a$.  

The photon asymmetry $\Sigma$ is given by
\begin{equation}
\Sigma = \frac {d \sigma _{\parallel} (\vec{\gamma},NN) -
                d \sigma _{\perp} (\vec{\gamma},NN) }
                {d \sigma _{\parallel} (\vec{\gamma},NN) +
                d \sigma _{\perp} (\vec{\gamma},NN) } \; ,
\end{equation}
where $d \sigma _{\parallel (\perp)}$ is the differential cross
section for photons linearly polarized parallel (perpendicular) to
the $xz$ reaction plane.  The polarization for the nucleon $N_a$ 
is defined according to
\begin{equation}
P_y^a = \frac {d \sigma ^{\uparrow} (\gamma,\vec{N}_aN) -
                d \sigma ^{\downarrow} (\gamma,\vec{N}_a N) }
                {d \sigma ^{\uparrow} (\gamma,\vec{N}_aN) +
                 d \sigma ^{\downarrow} (\gamma,\vec{N}_a N) } \; ,
\end{equation}
where $d \sigma ^{\uparrow (\downarrow)}$ is the differential cross
section for nucleon $N_a$ polarized along (against) the $y$-axis. 
Remark that the choice of the reference system is such that the
nucleon on which recoiling nucleon polarimetry is
performed is labelled as $N_a$ and that the spin is measured along the
direction normal to the reaction plane containing $\vec{q} _{\gamma}$ and
$\vec{k}_a$.  

The unpolarized differential cross section and asymmetry  
can be written in terms of structure functions and read \cite{boffi1}
\begin{eqnarray}
\frac {d^4 \sigma}
{d \Omega _a d \Omega _b dE_a dE_b}
& = & \frac {1} {(2\pi)^5 E_{\gamma}} k_a k_b E_b E_a 
\delta (E_{A-2} + E_a + E_b - E_A - E_{\gamma}) \frac {1} {2} W_T
\\
\Sigma & = & - \frac {W_{TT}} {W_{T}} \; ,
\label{eq:sigma}    
\end{eqnarray}
with
\begin{eqnarray}
W_T(q _{\gamma},k_a,k_b,\theta _a,\theta _b,\phi _a,\phi _b) & = & 
\sum_{m_{s_{a}},m_{s_{b}},M_R} \left[
\left( m_F^{fi}(\lambda=+1)\right)^*\left(m_F^{fi}(\lambda=+1)\right)
\right. \nonumber \\  
& & +
\left. \left(m_F^{fi}(\lambda=-1)\right)^*\left(m_F^{fi}(\lambda=-1)
\right) \right]
\label{wt}
\\ 
W_{TT} (q _{\gamma},k_a,k_b,\theta _a,\theta _b,\phi _a,\phi _b)& = & 2 Re \Biggl[
\sum_{m_{s_{a}},m_{s_{b}},M_R}
\left(m_F^{fi}(\lambda=-1)\right)^*
\left(m_F^{fi}(\lambda=+1)\right)
\Biggr] \;.
\label{wtt}
\end{eqnarray}
At this point it is worth remarking that the unpolarized $(e,e' N_a
N_b)$ cross section depends on both the $W_T$ and $W_{TT}$ structure
function.  Apart from these, the $(e,e' N_a N_b)$ differential cross
section is further determined by the $W_{LT}$ and $W_{L}$ structure
function,  
reflecting the longitudinal degree of freedom in electron scattering.
In any case, with the aid of polarized real photon beams one can
gain control over the  transverse channels which have been shown to
contribute substantially to triple
coincidence reactions of the $(e,e'N_aN_b)$ type \cite{giusti97,janeepp}.

The outgoing nucleon polarization can be written in the following form
\begin{eqnarray}
P_y^a= \frac {1} {W_T} \sum _{m_{s_{b}},M_R, \lambda \pm 1} Re \Biggl[
 & & i \left< \Psi_{f}^{(A-2)}(E_x,J_R M_R);
{\vec k}_a m_{s_{a}}=1/2 ;{\vec k}_b m_{s_{b}}
\mid J_{\lambda} ({\vec q}) \mid \Psi _0 \right> 
\nonumber \\
& & \times \left< \Psi_{f}^{(A-2)}(E_x,J_R M_R);
{\vec k}_a m_{s_{a}}=-1/2 ;{\vec k}_b m_{s_{b}}
\mid J_{\lambda} ({\vec q}) \mid \Psi _0 \right>^* \Biggr]   \; .
\end{eqnarray}

\section{Models for exclusive two-nucleon knockout from finite nuclei}
\label{sec:model}
Direct two-nucleon knockout following photoabsorption is dominated by
two-body current operators $J_{\lambda}^{[2]} ({\vec q})$.
In our model calculations, an effective Lagrangian approach is adopted
to construct the two-body currents that determine the coupling of the
photon field to the nuclear system.  In selecting the current
operators or, equivalently, the contributing reaction
mechanisms, we have been led by the observation 
that the dominant role of (virtual)
pion production in photoinduced two-nucleon knockout has been established
in several photoabsorption calculations
\cite{carrasco,wakamatsu,hutt}.  The most elementary process that
allows gaining control over the role of the pion in photoinduced experiments
is pion photoproduction on the nucleon $\gamma + N \longrightarrow N'
+ \pi$.  It is very well known that 
for photonenergies below 0.5~GeV the cross
sections and polarization observables for these reactions are 
fairly well understood in an effective Langrangian approach retaining a
selected class of diagrams \cite{ew,marcpion}.
The diagrams which are commonly included in photopion production studies 
are shown in
Fig.~\ref{fig:pion}.  We now transfer all elementary pion production
processes of Fig.~\ref{fig:pion} to the nuclear medium and
assume that the created pion is reabsorbed.  As a result of such a
process  a
second nucleon can be brought on the mass shell and one obtains the diagrams of
Fig.~\ref{fig:twon} all feeding the two-nucleon knockout channel.
These diagrams represent the processes which are believed to be the
major contributors to the $(\gamma,NN)$ reaction process at low
missing energies.  In Figure~\ref{fig:twon}, diagrams (a), (b), (d) and (e)
correspond with genuine two-body photoabsorption, respectively related
to meson-exchange ((a) and (b)) and isobaric currents ((d) and (e)).
Diagrams (c) and (f) are characterized by one-body photoabsorption
with subsequent two-nucleon knockout and refer to final-state
interactions and ground-state correlations respectively.

The two-body operators related to Fig.~\ref{fig:twon}(a)~and~(b) are
the standard pion-exchange current operators that can be derived from
the one-pion exchange part of the nucleon-nucleon interaction through
minimal substitution \cite{ew,riska}.  We
use the standard $\pi NN$ pseudo-vector coupling with $f_{\pi
  NN}^2/4 \pi$ = 0.079 and regularize the $\pi NN$ vertices with
monopole form factors using a cut-off mass of $\Lambda _{\pi
  NN}=1200~MeV$.  The one-nucleon photoabsorption mechanism of
Figure~\ref{fig:twon}(f) would not contribute in a shell-model (or
independent particle) picture of the target nucleus and refers to the
situation in which one-body photoabsorption on a correlated pair
forces two nucleons to escape.  In principle, this process could be
treated on the same footing as the other contributing diagrams.  This
would imply, however, that solely ground-state correlations induced by
the one-pion exchange part of the nucleon-nucleon interaction are
included.  This is at odds with the widely spread belief that heavier
meson exchange is the major source of ground-state correlations.
Generally, short-range effects are difficult to deal with in a
diagrammatic approach.  Nevertheless, various techniques have been
developed to cope with the complexity of the ground-state of an
interacting Fermion system.  Here, we resort to a semi-phenomenological
approach to relate the results of these theoretical 
ground-state investigations with the cross
sections and polarization observables for
$(\gamma,NN)$ reactions.  It consists of correcting the mean-field
Slater determinant  $\mid \Psi _o >$ with a Jastrow-like correlation
function :
\begin{equation}
\prod _{i<j} f_c (r_{ij}) \mid \Psi _o > \; .
\end{equation}
Such an expansion mocks up two-, three-, \ldots A-body correlations.
In the actual calculations only the central correlations of the two-body type
are retained.  As a matter of fact, such an approximation amounts to correcting
the product of the single-particle bound-state functions of the active
nucleon-pair $\phi _a (1) \phi _b (2)$ with a central 
correlation function :
\begin{equation}
\phi _a (1) \phi _b (2) \longrightarrow \phi _a (1) \phi _b (2)
f_c(r_{12}) \; .
\end{equation}
In an attempt to find out about the sensitivity of the $(\gamma,NN)$ 
angular cross sections and
polarization observables to the ground-state correlations we have used
several types of correlation functions in the calculations.   

The results of this paper will confirm the importance of the $\Delta
_{33}$ resonance in photoinduced two-nucleon knockout for
photonenergies below 0.5~GeV.  As we consider
photon energies in the resonance region special care must be taken in
constructing the isobaric current operators related to intermediate
$\Delta _{33}$ creation of the type $\gamma NN
\longrightarrow N \Delta \longrightarrow NN$.  A detailed account of
the isobaric currents in the context of two-nucleon knockout reactions
was given in Ref.~\cite{janeepp}.  As our model calculations start
from the assumption that the pion is the major mediator between
several nucleons, the $V_{NN,N \Delta}$ interaction which enters in
the construction of the isobaric currents is the pion exchange
interaction.

\subsection{A factorized model for photoinduced two-proton knockout}
\label{sec:fac}
We now consider a simplified model for $(\gamma,N_a N_b)$.  
Two approximations are necessary for the angular ($\gamma,NN$) cross
section to factorize.  First, a plane wave description for the
escaping particles should be adopted.  In addition, a severe
assumption regarding the relative motion of the pairs has to be
made in the sense that the decoupling of the c.m. and relative
motion of the pair can formally only be achieved after the photon is assumed to
couple to dinucleons residing in relative $S$ waves.  This
simplification is known as the ``quasi-deuteron'' approximation.  This
terminology might be a bit confusing in the sense that is used for both
proton-proton and proton-neutron knockout.  Furthermore, the
range of the relative $S$ wave is assumed to be small in comparison
with all other mechanisms that play a role in the reaction process.
From the technical point of view, this allows replacing the relative
$S$ wave by a delta function (known as the ``zero-range
approximation''), thus enormously simplifying the
theoretical calculations \cite{boato,janqd,janplb,marcvdh}.   
The ``zero-range approximation'' somehow implies that one simplifies 
the Fermi motion in the relative wave function of the pair.
The factorized scheme can be equally well applied to $(\gamma,pn)$ and 
$(\gamma,pp)$ reactions.  The fact that solely the terms related to
the ground-state correlations and a few pieces from the isobaric
current contribute to the direct proton-proton knockout mechanism
makes the factorized model to be particularly attractive as rather simple
analytical expressions for the asymmetry and cross sections can be
derived.  The procedure for arriving at factorized expressions for the
structure functions was outlined in Ref.~\cite{janplb}. 
For the sake of completeness we rewrite
the derived expressions for the $W_T$ and $W_{TT}$ structure functions
at the real photon point 
\begin{eqnarray}
W_T & = & F_{hh'}(P) \Biggl\{  \frac {\mu _p ^2 e^2 q^2_{\gamma}} {M_p^2} 
\left( g(k_+) - g(k_-) \right)^2  \nonumber \\
   & + & \frac {e^2} {2M_p^2} 
     \left[
          \left( k_{a,x} g(k_-) + k_{b,x} g(k_+) \right)^2
        + \left( k_{a,y} g(k_-) + k_{b,y} g(k_+) \right)^2 
     \right]
\nonumber \\
& + &  \frac {256} {81} \left( \frac 
         {f_{\gamma N \Delta} f_{\pi N \Delta} f_{\pi NN}}    
         { m_{\pi} ^3}  \right)^2 
         G_{\Delta} ^2 
         \left( \vec{q} _{\gamma} \times \left( \frac
                                     {\vec{k}_a - \vec{k} _b}
                                     {2}
                          \right) \right)^2  
\times \left[ \vec{k}_+  \frac {1} {k_+^2+m_{\pi}^2} 
- \vec{k}_-  \frac {1} {k_-^2+m_{\pi}^2} \right]^2 \Biggr\} \nonumber \\
W_{TT} & = & F_{hh'}(P) \Biggl\{ \frac {- 2 \mu _p ^2 e^2 q_{\gamma}^2} {M_p^2} 
\left( g(k_+) - g(k_-) \right)^2 
\nonumber \\
   &   - & \frac {e^2} {M_p^2} 
     \left[
          \left( k_{a,x} g(k_-) + k_{b,x} g(k_+) \right)^2
        - \left( k_{a,y} g(k_-) + k_{b,y} g(k_+) \right)^2 
     \right]
\nonumber \\
& - &  \frac {256} {81} \left( \frac 
         {f_{\gamma N \Delta} f_{\pi N \Delta} f_{\pi NN}}    
         { m_{\pi} ^3}  \right)^2 
         G_{\Delta} ^2  
         \left[ \left( \vec{q} _{\gamma} \times \left( \frac
                                     {\vec{k}_a - \vec{k} _b}
                                     {2}
                          \right) \right)_x ^2  
- \left( \vec{q} _{\gamma} \times \left( \frac
                                     {\vec{k}_a - \vec{k} _b}
                                     {2}
                          \right) \right)_y ^2 \right]
                             \nonumber \\
& & \times \left[ \vec{k}_+  \frac {1} {k_+^2+m_{\pi}^2} 
- \vec{k}_-  \frac {1} {k_-^2+m_{\pi}^2} \right]^2 \Biggr\} \; ,
\label{eq:wtwtt}
\end{eqnarray}
with
\begin{equation}
\vec{k}_{\pm} = \frac {\vec{k}_{a} -\vec{k}_{b}} {2}  
\pm \frac {\vec{q} _{\gamma}} {2}  \nonumber \; .
\end{equation}
The function $g(k)$ appearing in the above expression is the Fourier
transform of the central correlation function
\begin{equation}
g(k)  \equiv  
\int d \vec{r} e ^{i \vec{k}. \vec{r}} (1-f_c(r)) \; ,
\end{equation} 
and $F_{hh'}(P)$ is the probability of finding a proton pair with
quantum numbers $(h(n_h l_h j_h),h'(n_h l_h j_h))$, c.m. momentum
$P$ and small internucleon separations in the target nucleus.  
The $G_{\Delta}$ is the sum of the
``resonant'' (direct $\Delta$ term corresponding with Fig.\ref{fig:twon}(d))  
and non-resonant $\Delta$ propagator
(crossed  $\Delta$ term corresponding with Fig.\ref{fig:twon}(e)). 
\begin{eqnarray}
G_{\Delta} & \equiv & G_{\Delta}^{res} + G_{\Delta}^{non-res}
\nonumber \\ & = & {{1} 
                    \over
                    {-E_{\Delta}^{res}+M_{\Delta} - {i \over 2} \Gamma
                     _{\Delta}^{res}  + V_{\Delta} }} 
                    + {{1} \over
                    {-E_{\Delta}^{non-res}+M_{\Delta} }} \; ,
\label{eq:propa}
\end{eqnarray}
where $M_{\Delta}=~$1232~MeV, $\Gamma_{\Delta}^{res}$ is the free $\pi
N$ decay width and 
$V _ {\Delta}$ is the $\Delta$ selfenergy in the medium.  For light
nuclei the imaginary part of this selfenergy is established to be
40~MeV  at full nuclear density \cite{ew,salcedo}.  The $E_{\Delta}^{res}$ 
and $E_{\Delta}^{non-res}$ are the invariant energies of the $\Delta$ 
in respectively the diagram of Fig.~\ref{fig:twon}(d) and (e). 
In deriving the analytical expressions of 
Eq.(\ref{eq:wtwtt}) the $\Delta$-current terms in 
($G_{\Delta}^{res} - G_{\Delta}^{non-res}$) have been neglected.     

The first two terms in the expressions for $W_T$ and $W_{TT}$ 
 refer to one-body
photoabsorption on the magnetization and convection current
respectively. Retaining only the contribution from $\Delta$ current
to the above structure functions and using the indentity
(\ref{eq:sigma}) one readily obtains the following expression
for the $(\gamma,pp)$ asymmetry 
\begin{equation} 
\Sigma = - \frac {k_a^2 sin ^ 2 \theta_a cos 2 \phi _a
+ k_b^2 sin ^ 2 \theta_b cos 2 \phi _b
-2 k_a k_b sin \theta _a sin \theta _b cos (\phi_a + \phi_b)}
{k_a^2 sin ^ 2 \theta_a 
+ k_b^2 sin ^ 2 \theta_b 
-2 k_a k_b sin \theta _a sin \theta _b cos (\phi_a - \phi_b)} \; ,
\end{equation}
where ($\theta$,$\phi$) are the polar and azimuthal angle of the
escaping nucleons.
For planar kinematics the isobar contribution to $W_T$ and $W_{TT}$ is
equal and in the absence of ground-state correlations the factorized
model predicts that the $(\vec{\gamma},pp)$ asymmetry is exactly -1.
At first sight this conclusion seems to be at odds with the fact
that symmetry observations require $\Sigma=0$ for $\theta_1,\theta_2
\epsilon \{0^o,180^o\}$.  Under these circumstances, however, also the
$\Delta$ current contributions to the $W_T$ and $W_{TT}$ structure
functions are predicted to be exactly zero for both photon
polarizations, leaving the photon asymmetry $\Sigma$ undetermined.  
Remark that the
major contribution from the ground-state correlations, namely the one
originating from photoabsorption on the one-body 
magnetization current \cite{janplb}, does equally contribute to
$W_T$ and $W_{TT}$.  Moreover, in both cases it has the same relative
sign with respect to the isobaric terms.  For that reason the
factorized scheme suggests that no particular sensitivity of the
$(\gamma,pp)$ asymmetries to ground-state correlations should be
expected.  The above expression further suggests a sort of universal
behaviour for the asymmetry of direct $(\gamma,pp)$ processes, which
is independent from the shell-model structure of the active pair or
target mass number A and only involves the momenta of the two ejected
protons.  We did not succeed in deriving a simple form for the
$P_y^p$.  In order to reach scaling in terms of the pair function
$F(P)$ one relies on closure properties and such a procedure seems to
be excluded when nucleon spin observables are probed.

\subsection{Unfactorized model for photoinduced $2N$ knockout}
\label{sec:unfac}
In Refs.~\cite{gent,janeepp} an unfactorized model for calculating
two-nucleon knockout cross sections was presented.  In comparison with
the factorized model outlined in previous Subsection it provides a
more realistic description of the escaping nucleon wave functions.
Moreover, no assumptions are made regarding the nature of the relative
wave function of the active nucleon pair.  As a matter of fact, the
unfactorized model does not make the separation between relative and
center-off-mass motion as this can only be achieved in a harmonic
oscillator basis.  Instead of working in a harmonic oscillator basis, we use
mean-field quantities (wave functions, potentials and phase shifts)
from a Hartree-Fock calculation with the effective Skyrme force SkE2
\cite{waroquier}.  We deem these single-particle wave functions to be
realistic, as we could obtain a good description for the exclusive
quasi-elastic $(e,e'p)$ cross sections using these wave functions and
modern optical potentials \cite{veerleeep}.  Here, we solely summarize
the physical ideas behind the unfactorized model and refer to
Refs.~\cite{gent,janeepp} for the technical details.  The principal
idea is that following the absorption of a photon by the target nucleus,
two nucleons are excited from a bound into a continuum eigenstate of a
mean-field potential.  By making a proper partial wave expansion with
products of these continuum eigenstates, an antisymmetrized wave
function can be constructed that is characterized by two
asymptotically escaping nucleons and a residual $A-2$ fragment.  In
order to ensure the antisymmetrization of the final wave function we
found it appropriate to start from a basis set in which the residual
fragment is created with a particular angular momentum $J_R$.  Within
the outlined approach, the essential quantities that enter the
calculation of the two-nucleon knockout cross sections are the reduced
transition matrix elements of the type
\begin{equation}
<p(\epsilon lj) p'(\epsilon' l'j');J_1 \| T_{J} (q_{\gamma}) 
\|h (n_h l_h j_h) h' (n_{h'} l_{h'} j_{h'}) ; J_R > \; ,
\label{eq:matrixel}
\end{equation}
where $h$ ($h'$) are the bound state wave functions for the orbits
from which the nucleons are escaping and $p$ ($p'$) the continuum
eigenfunctions of the mean-field potential.  The $T_J$ stands for the
magnetic and electric transition operator which is obtained after
making a multipole decomposition of the current operators.  For the
results presented here we reached convergence after including all
electric and magnetic multipoles up to $J$=5.  Each of the diagrams of
Figure~\ref{fig:twon} will contribute to the transition operator
$T_J$.  For the meson-exchange and isobaric currents the expressions
for the matrix elements are given in the Appendix of Ref.~\cite{gent}.
It should be noted that in the latter reference, which concentrated on
photonenergies below the resonance region, the static limit for the
$\Delta$ propagators were considered.  Here an updated version that
uses dynamic $\Delta$ propagators and 
accounts for both the direct and crossed $\Delta$ terms has been used.
The corresponding current operator was described in detail in 
Refs.~\cite{janeepp,marcvdh}.  This current operator could fairly well 
describe the photon energy dependence of the $^{12}$C($\gamma,pn$) and 
$^{12}$C($\gamma,pp$) cross sections through the $\Delta$ resonance region
\cite{douglas}.  
The procedure for including the ground-state correlation effects
of Figure~\ref{fig:twon}(f) was outlined in Ref.~\cite{janeepp},
where also the expression for the corresponding 
reduced matrix elements are given.    We would like to stress  
that the unfactorized model can be used with plane wave outgoing
nucleon waves.  It suffices to use Bessel functions for the continuum
single-particle states $p(\epsilon lj)$ and $p'(\epsilon ' l'j')$ in the  
above matrix element.  The fact that the same Hamiltonian is used to
construct the mean-field and the bound-state wave functions, together
with the transverse nature of the isobaric current, makes the
$(\gamma,pp)$ calculations gauge invariant.

\section{Results and discussion}
\label{sec:results}
In what follows, predictions for $^{16}$O and $^{12}$C($\gamma$,$NN$)
differential cross sections and polarization observables will be presented.
The sensitivity of the measurable quantities to the various aspects of
the reaction mechanism will be discussed.  The role of the final state
interaction can be assessed by considering that the unfactorized model
can be applied with either distorted or plane wave outgoing nucleon
wave functions, keeping all other ingredients of the calculations
exactly equal.  The extent to which proton-proton knockout is
dominated by $S$ wave absorption, will be estimated by comparing the
results obtained within the context of the full unfactorized model
with the predictions from the
factorized scheme.  Indeed, the latter model is explicitly based on
the dominance of $^1S_0$ proton pairs whereas no such restrictions are
made for the unfactorized approach.  In the course of this section we
will suggest another way of gaining insight into the role of the
different combinations for the 
relative wave function  in the photoabsorption mechanism.
The method is based on the
selective nature of the photoabsorption mechanism when the final state
is created with a specific angular momentum $J_R$.

The reason for concentrating on p-shell nuclei is partly inspired by
the observation that the bulk of the 
experimental activities are concentrated in this part of the mass
table.  For the heavier target nuclei the transparency, which is
related to the probability for 
nucleons to escape, is considerably smaller \cite{garino}. 
 Accordingly, the ambiguities with respect to the FSI in two-nucleon 
emission reactions are expected to become increasingly worrying 
with increasing target mass number A.

We consider two types of
kinematical conditions that create favourable conditions to study the
dynamics of pairs in the medium and present exclusive cross sections
for (groups of) states in the low missing-energy range.

\subsection{$^{16}$O($\gamma,NN$) and $^{12}$C($\gamma,NN$) 
in quasi-deuteron kinematics}

In this Subsection we are presenting cross sections 
in so-called ``quasi-deuteron''
(QD) kinematics which is constrained by imposing the
condition that the missing momentum $\mid \vec {P} \mid$ equals zero.
Accordingly, one is considering photoabsorption on dinucleons which
are at rest (this of
course in an ideal world in which final state interactions could be
completely ignored).  
For in-plane
kinematics, the QD condition leads to a unique solution for
$\theta _b$, $\mid \vec {k} _a
\mid$ and $\mid \vec {k} _b \mid$ for each given set of ($\vec{q} _{\gamma}$,
$\theta _a$, $E_x$).  The effect of the c.m.  motion 
 on the angular 
cross sections is then minimized as the pair function $F(P=0)$ remains
constant.  Accordingly it is hoped that a maximized sensitivity
to the dynamics of the pair relative motion can be achieved.  
Quasi-deuteron kinematics further 
allows comparing A$(\gamma,pn)$ observables with
results obtained in photodisintegration of the  deuteron.
This opens perspectives to study in how far proton-neutron pairs in
the medium resemble quasi-deuteron properties and to learn
more about possible medium modifications of 
the photon-nucleus coupling.  
From the experimental point of view 
QD kinematics has the obvious advantage of sampling that
region of the phase space for which the largest cross sections
are expected.

It is worth reminding that we have used realistic
single-particle wave functions in our calculations and do not rely on
an expansion in terms of relative and c.m. Harmonic-Oscillator wave
functions in order to calculate the cross sections.  Nevertheless, it
turns out that considering the different relative and c.m.
combinations in the Harmonic Oscillator limit is very helpful in
interpreting the general behaviour of the angular cross sections and
polarization observables.   
In Tables \ref{tab:pp} and \ref{tab:pn} we have collected all the
possible quantumnumbers of the pair relative and c.m. wave functions
for proton-proton and proton-neutron knockout from the different s-
and p-shell combinations. The combinations are made as if the bound
state wave functions are of the simple Harmonic-Oscillator type.
After making a separation in the relative and c.m. motion through a
Moshinsky transformation the antisymmetric two-body wave functions
read  
\begin{eqnarray}
& & 
\left| 
\left( n_a l_a j_a t_a, n_b l_b j_b t_b \right) ;  J_R M_R \right>_{as} 
  =    
\sum _ {L M_L} \sum _{n l} \sum _{N \Lambda} 
\sum _{S M_S} \sum_{T M_T} 
\widehat{j_a}   \widehat{j_b} \widehat{L} \widehat{S} 
 \left< \frac{1}{2} \; t_a \; \frac{1}{2} \; t_b \mid T \; M_T \right>
\nonumber \\ 
& &    \times \left<L \; M_L \; S \; M_S \mid J_R  \; M_R   \right>
\left\{ \begin{array}{lll}
    l_a \; l_b \; L    \\ 
    \frac{1}{2} \; \frac{1}{2} \; S  \\ 
    j_a \; j_b \; J_R
    \end{array} \right\} 
 \left<nl,N\Lambda;L \mid n_a l_a,n_b l_b;L \right>  \nonumber \\ 
& & \times \left|(nl,N\Lambda)LM_L,\left(\frac{1}{2} \frac{1}{2}\right)SM_S, 
\left(\frac{1}{2} \frac{1}{2}\right)TM_T \right> 
\left[1-(-1)^{l+S+T}\right] \; ,
\end{eqnarray}
where $\widehat{j} \equiv \sqrt{2j+1}$, $T$ ($S$) is the total isospin 
(spin) of the pair and $l$
($\Lambda$) the angular momentum of the relative (c.m.) pair wave
function.  In the above expression we have used the conventions of
Ref.~\cite{heyde} for the angular momentum coupling coefficients. 
For direct two-nucleon knockout from a $0^+$ target
nucleus the {\em total} angular momentum of the pair
 (which is the sum of the c.m. angular momentum, relative angular
momentum and total  spin of the pair) 
determines the quantum number $J_R$ of the residual fragment.
By studying the cross section for the different 
$J_R$'s, which could experimentally be achieved in a high-resolution
experiment \cite{gerco,lennart}, one can then study the behaviour of the 
observables as a
function of the various types of relative pair wave functions.  In QD
kinematics, pair c.m. wavefunctions with an angular momentum $\Lambda
> 0$ are unlikely to contribute substantially to the differential
cross sections.  Assuming that the inital photoabsorption occurs on a
relative $S$ state, then knockout from the (1p$_{3/2}$)$^2$
configuration is expected to populate a $J_R=0^+$ (proton-proton case)
and $J_R=1^+$ (proton-neutron case).  We have calculated the
observables for proton-proton and proton-neutron knockout from
$^{12}$C considering QD kinematics and four representative photon
energies.  The results are summarized in Figs.~\ref{fig:seapp} and
\ref{fig:seapn} and refer to the situation in which the $A-2$ fragments
$^{10}$B and $^{10}$Be are created in a (1p$_{3/2}$)$^{-2}$ two-hole
state.  The figures show the contribution for the individual angular
momentum components  $J_R$ of the A-2 fragment as well as the result (solid lines)
which is obtained after incoherently adding all the components.
Accordingly, the solid lines reflect the situation in which an
integration over the missing-energy range for p-shell knockout would
be made.  

The results of Figs.~\ref{fig:seapp} and \ref{fig:seapn} are in
conformity with the above predictions in the sense that the major
contribution to the respective cross sections comes indeed from
$J_R$=0$^+$ (proton-proton case) and $J_R$=1$^+$ (proton-neutron
case).  These are the states that one would expect to be populated
assuming absorption on dinucleons in relative $S$ waves and
c.m. angular momentum $\Lambda$=0.  The latter type of components are 
naturally favoured in QD kinematics.  Nevertheless, the calculations
produce substantial contributions from configurations that fall
beyound this QD picture.
The population of the $0^+$ state in the $(\gamma,pn)$ reaction 
with a non-negligible cross section
could be interpreted as 
a manifestation of photoabsorption on $^1S_0(T)$=1 proton-neutron pairs which
is a slightly unbound configuration in the free proton-neutron system.
The next two important $J_R$'s contributing to the $(\gamma,pn)$ 
observables are $J_R=0^+$ and $3^+$.
The $J_R=2^+$ cross section is very small.    
It should be stressed that all above statements regarding the
relative population of the states apply to QD kinematics and should
not be considered as general.  When
moderate and larger values of the missing momentum $P$ are probed,
configurations with c.m. angular momenta $\Lambda > 0$ are expected to start 
playing a major role
and the relative population of the different angular momentum states
of the $A-2$ fragment is expected to be different.  Moving out of QD
kinematics is expected to make the role of the c.m. 
 configurations with $\Lambda > 0 $more important.  At the same time, the cross
sections will become smaller as larger values of the missing momentum
are probed.    

When comparing the corresponding proton-proton and proton-neutron
results of Figs.~\ref{fig:seapp} and ~\ref{fig:seapn} one observes
that the $(\gamma,pn)$ asymmetries are generally smaller than the
$(\gamma,pp)$ ones. A similar remark holds for the polarizations.
It is also worth remarking that the shapes of the proton-proton and
proton-neutron differential cross sections are considerably
different at corresponding photon energies.  The angular dependency
and magnitude of the polarization observables is noticed to be
remarkably close to the ones for the $J_R=0^+$ ($(\gamma,pp)$) and
$J_R=1^+$ ($(\gamma,pn)$) component.  Even more than the cross sections
the polarization observables seem to be dominated by photoabsorption on
$(l=0,\Lambda=0)$ pairs in QD kinematics.  This seems to be
particularly the case for the higher end of the photon energies
considered here and is more pronounced in the $(\gamma,pn)$ channel. In
an attempt to investigate whether $S$ wave absorption automatically
implies that the ($\gamma,pn$) observables exhibit a deuteron
like behaviour we have compared the $^{12}$C($\gamma,pn$) asymmetries
with d($\vec{\gamma},p$)n data at corresponding photon energies.  At
E$_{\gamma}$=100~MeV the deuteron asymmetry has  very little
resemblance with the $^{12}$C predictions.  This points towards
different underlying reaction mechanisms.  Whereas, one-body
photoabsorption plays an important role for the d$(\gamma,p)$n observables
at low photon energies \cite{fritz,levchuk}, 
the calculations predict that $^{12}$C($\gamma$,pn) is
dominated by pion exchange currents.  As one moves into
the $\Delta$ resonance region, the calculated $^{12}$C asymmetries move
closer to the deuteron data.  At  E$_{\gamma}$=300 and 400~MeV the
agreement of the $^{12}$C asymmetries with the deuteron data is even
remarkable.  At these energies, also the $^{12}$C proton polarizations
$P_y^p$  turn
out to have the same sign and similar magnitude than what was obtained in
d($\gamma,\vec{p}$)n measurements.   

We now invesigate the sensitivity of the ($\gamma,pn$) observables to
the different ingredients which enter the calculations.
Fig.~\ref{fig:figa} shows the calculated $^{16}$O($\gamma,pn$) results
for E$_{\gamma}$=150~MeV in QD kinematics.  We consider the situation
whereby the two nucleons are emitted
from the 1p shell.  Two different types of shell-model configurations
were considered: ($1p_{3/2}$)$^{-2}$ and
($1p_{3/2}$)$^{-1}$($1p_{1/2}$)$^{-1}$.  To obtain the curves of
Fig.~\ref{fig:figa} the contributions from the $J_R$
components to the various cross sections ($\sigma,
 \sigma ^{\uparrow},  \sigma ^{\downarrow},  \sigma _{\parallel}, 
\sigma _{\perp}$) were
incoherently added.   The sensitivity to the
various terms in the photoabsorption process is illustrated by
comparing the dashed and dotted line.  Relative to a calculation that
solely accounts for the seagull pion-exchange diagram
(Fig.~\ref{fig:twon}(a)), the inclusion of the pion-in-flight term
(Fig.~\ref{fig:twon}(b)) reduces the cross section and does even
switch the sign of the asymmetry. This illustrates the sensitivity of
the latter to the different contributing terms in the pion exchange
part of the photoabsorption
mechanism.  Remark that even at photonenergies as low as 150~MeV the effects
from the isobaric current are sizeable.  The dot-dashed line uses the
full photoabsorption operator but plane waves for the escaping proton
and neutron wave function.  Comparing these results with the solid
line, which is obtained with exactly the same current operator but
using distorted outgoing nucleon waves, one can estimate the role of
the final-state interaction.  The asymmetry is hardly affected by the
outgoing nucleon distortions, whereas the differential cross section
is roughly reduced by a factor of two.  The effect of the distortions
on the proton polarizations is large.  Its magnitude is of the same order
as the deuteron results at corresponding photon energies
\cite{arenhovel,fritz,levchuk}.  In comparing the $(\gamma,pn)$ results of
Fig.~\ref{fig:figa} for both shell-model configurations one gets a
feeling about the nuclear structure dependency of the cross sections
and polarization observables.  In this context it is worth remarking
that in the Hartree-Fock basis that we are using 
the radial dependence of the  $1p_{3/2}$ and $1p_{1/2}$ 
single-particle wave functions differs.  The
general trends for all observables are rather uniform for both
configurations.  The subtle differences between the two
situations are mainly a manifestation
for the importance of mechanisms
that go beyond $^3S_1$ absorption.  Indeed if there would be solely
quasi-deuteron like absorption on ($^3S_1$(T=0),$\Lambda$=0) pairs both
shell-model configurations would be dominated by the $J_R$=1$^+$
contribution.

Predictions for the $^{12}$C$(\gamma,pp)$ cross section within the
simple factorized model are shown in Fig.~\ref{fig:overpp} and are compared
with the full (unfactorized) model calculations.  At low photon
energies, where the range of the photons is really too large to expect
mere photoabsorption on short-ranged $S$-waves, the factorized model largely
undershoots the full model calculations.  With increasing photon
energy, the factorized predictions seem to get closer to the full
model predictions.  The full model calculations, however, produce
differential cross sections which are not as sharply peaked.
Referring to the results of Fig.~\ref{fig:seapp}, which shows the
different $J_R$ contributions for the distorted wave (solid line)
calculation of Fig.~\ref{fig:overpp}, this is mainly due to excitation
of the $J_R$=1$^+$ state, which is somehow excluded when considering
($^1S_0$,$\Lambda$=0) ``quasi-deuteron''  absorption.  Remark further
that the proton angle dependence 
for the $J_R$=0$^+$ contribution in the unfactorized model (Fig.~\ref{fig:seapp}),
which is the configuration that dominant S-wave photoabsorption would
select, bears a strong resemblance with the  factorized model's
prediction.  As a consequence, it can be concluded that a major source
of deviations between the results produced by the factorized scheme
and the full model calculations has to be ascribed to photoabsorption
mechanisms that go beyond a quasi-deuteron like mechanism.  The effect
of the final state interaction can be estimated by comparing the solid
and dashed line in Fig.~\ref{fig:overpp}.  As is usually the case, the
final-state distortions tend to widen the peaks of the angular cross
sections  obtained in the plane-wave approximation.  
The effect of the outgoing nucleon
distortion is particularly large for the left panel of
Fig.~\ref{fig:overpp}.  The typical outgoing nucleon kinetic energy is
20-50~MeV, which makes this result not that surprising.  Remark that
for this low photonenergy the effect of the FSI on the $\Sigma$ is
even sizeable.  For the right panel, corresponding with outgoing
nucleon kinetic energies ranging from 60-210~MeV, the role of the
outgoing nucleon distortions is minor but not negligible.

A plane wave calculation with a Hermitian current operator would
produce an outgoing nucleon polarization which is exactly zero.  Two 
mechanisms of completely different origin 
can make $P_y^p$ different from zero.  First, final state
interaction effects and secondly, the fact that the $\Delta$
propagator $G_{\Delta}^{res}$ of Eq.~(\ref{eq:propa}) 
entering the isobaric current, contains
imaginary parts.  At low photon energies, where the damping effects of
the $\Delta$ propagator can only play a marginal role, one expects a
plane wave calculation to produce $P_y^p$=0.  This fundamental property
can be exploited to check the numerical accuracy of the
calculations. It speaks in favour of the calculations that the plane
wave result at $E_{\gamma}$=100~MeV does indeed obey this criterion
for the full range of proton emission angles.  At $E_{\gamma}$=100~MeV, the
polarization is thus fully determined by the FSI mechanisms.
In the resonance region, the occurrence of the
$\Delta$ width would make a plane wave calculation already produce a
$P_y^p \neq$0.  The $P_y^p$ is then a measure for the medium-dependent 
damping mechanism of  
intermediate $\Delta$ creation and the FSI effects. 
This is confirmed by the results in the
right panel of Fig.~\ref{fig:overpp}, where neither of the two
mechanisms is observed to be dominant.

Now we address the question in how far $(\gamma,pp)$ polarization
observables could be helpful in discriminating between the different
model predictions for the ground-state correlation effects.  Within
the context of the present model this amounts to investigating the
sensitivity to the different choices for the central correlation
function (sometimes referred to as  defect function).  In order to minimize
the contributions from the isobaric currents, we have considered
quasi-deuteron kinematics and photon energies on either side of the
$\Delta _{33}$ resonance : (a) E$_{\gamma}$=100~MeV and (b)
E$_{\gamma}$=400~MeV.  Referring to Eq.~(\ref{eq:wtwtt}) the effect of
the ground-state correlations, which primarily comes through the
magnetization current, is predicted to have a $q _ {\gamma} ^2$ dependence,
which make the SRC effects more likely to manifest themselves
at higher photon energies.  The left panel of Fig.~\ref{fig:wasrc},
which refers to E$_{\gamma}$=100~MeV, illustrates that even at lower
photon energies the ground-state correlation effects do not overshoot
the (suppressed) contribution from isobaric currents and exception
made for the hard core correlation function of Ref.~\cite{omy} 
(dot-dashed line)
which is generally considered not to be very realistic.  As a matter
of fact, it is obvious from Fig.~\ref{fig:wasrc} that dedicated
($\gamma,pp$) 
experiments could unambiguously settle the unrealistic character of
hard-core correlation functions.  Remark that there are
relatively little uncertainties regarding the $\Delta$ current
operator at low photon energies as the static limit for the isobaric
current operator would be fully justified.
Using realistic soft-core correlation functions, the ground-state
correlation effects are relatively more visible in the  
differential cross section and proton polarization than in the 
 photon asymmetry.  At E$_{\gamma}$=400~MeV the
strength from the isobaric currents remains sizeable and the effect of
the Jastrow correlations  on
the observables is of the
same size than for $E_{\gamma}$=100 MeV.  The results of
Fig.~\ref{fig:wasrc} illustrate that the mere fact of having the
photon absorbed on a $^1S_0$ diproton pair does not guarantee that the
reaction process is dominated by ground-state correlations.


\subsection{$^{16}$O($\gamma,NN$) in coplanar and symmetrical
kinematics}

Coplanar and symmetrical kinematics refer to a specific
situation for which the following kinematical conditions are obeyed~: ($ \mid
\vec{k} _a \mid $ = $ \mid \vec{k} _b \mid \equiv k$, $ \theta_a =
\theta _b = \theta$, $\phi_a = 0 ^o$, $\phi_b = 180 ^o$).  For each nucleon
emission 
angle $\theta$ and photonenergy, energy-momentum conservation will
provide a unique solution for the momentum $k$.  In the considered
kinematics the c.m. momentum P varies rapidly with the polar angle
$\theta _a$ and within the factorized model of Subsect.~\ref{sec:fac} 
we have $ \mid \vec{k} _+ \mid $ = $ \mid \vec{k} _- \mid =
\sqrt {k^2 sin ^2 \theta + q^2/4}$.  So, in many respects coplanar and
symmetrical kinematics is complementary to QD kinematics.  Whereas QD
kinematics would rather be used to compare the relative motion of
bound pairs 
with the deuteron, coplanar and symmetrical
kinematics is intrinsically meant to 
explore the fact that in the medium
nucleon pairs are characterized by c.m. degrees of freedom
besides their relative motion. The strong dependence of the cross
sections on the pair function $F(P)$, makes the absolute
magnitude of the cross sections to vary dramatically as a function of
the escaping nucleon  angle in coplanar and symmetrical kinematics.

Within the factorized $(\gamma,pp)$ model of Subsection~\ref{sec:fac},
the major contribution from the central correlations to the cross
section, which comes from a coupling of the photon field to the
one-body magnetization current, is exactly zero for $ \mid \vec{k} _+
\mid $ = $ \mid \vec{k} _- \mid $.  Accordingly, an abstraction made
of the (small) contribution from the convection current, both
the $W_T$ and $W_{TT}$  structure functions from Eq.~(\ref{eq:wtwtt}) are
predicted to be dominated by the isobaric current when facing coplanar and
symmetrical kinematics.

Figure \ref{fig:sunday} displays the 
calculated $^{16}$O$(\gamma,pp)$ and $^{16}$O$(\gamma,pn)$ cross
sections and photon asymmetries versus the proton emission angle
and photon energy. We have considered a (1p$_{1/2}$)$^{-2}$
configuration for the residual A-2 nucleus, so that only
$J_R$=0$^+$ can contribute for the $pp$ case, whereas the $pn$ channel
can feed two angular momenta states, namely $J_R$=0$^+$ and 1$^+$.  For both
channels and all emission angles considered 
a clear resonance at $E_{\gamma} \approx
$260~MeV is observed for the cross section.  The resonance is
not very visible in the asymmetry.  The $pn$ channel is characterized by
a wider photon energy and emission angle dependence than the $pp$
breakup channel.   In both cases, however, the pair function $F(P)$
makes the strength to reside in these parts of the phase space for
which the missing momentum $P$ is small.  At a fixed photon energy, 
the peak in the differential cross sections, which
corresponds with the situation that the missing momentum is
approximately zero, is localized at 
proton emission angles around 60-70 degrees in the lab frame.
Remark that $\theta _p
\approx 70^o$ corresponds with zero missing momentum and 
the maximum of the cross section  for the
low end of the photon energies considered here.  At higher photon energies
(E$_{\gamma} \geq$ 300~MeV) the peak does gradually shift to 60
degrees in the lab frame. The $P=0$ situation in coplanar and
symmetrical kinematics obeys also the QD conditions.

From the right panel of Fig.~\ref{fig:sunday} it becomes clear that
the $(\gamma,pn)$ asymmetry is characterized by rather strong
variations in both the photon energy and emission angle dependence.
The smallest values are reached for the low photon energies. With
increasing $E_{\gamma}$ the isobaric currents gain in relative
importance and as was already observed for the $(\gamma,pn)$ results
of Fig.~\ref{fig:figa} they tend to increase the asymmetry relative to
the values that one would get including solely the meson-exchange
contributions.  The attention is drawn to the fact that the
$(\gamma,pp)$ results of the left panel in Fig.~\ref{fig:sunday}
represent a rather unique case in the sense that the selected
(1p$_{1/2}$)$^{-2}$ final state can only feed the J$_R$=0$^+$ state
which gets its dominant contribution through ($^1S_0$,$\Lambda$=0)
photoabsorption.  An exception made for the lower photon energies, the
proton-proton asymmetry is close to -1 for the full range of
(E$_{\gamma}$,$\theta_p$) covered in Fig.~\ref{fig:sunday}.  We remind
that the asymmetry would be exactly -1 within the  factorized model of
Subsect.~\ref{sec:fac}.  This is another indication for the predictive
power of the factorized model when $^1S_0$ absorption is the dominant
configuration.
Our findings also confirm the calculations of Ref.~\cite{sandorfi},
where it was pointed out that even after including the c.m. motion of
the pair, simple $^1S_0$ diproton breakup produces
$^3$He($\vec{\gamma},pp$) asymmetries that are large and negative.

With the aim of studying in greater detail the impact of the different
relative angular momentum states on the $(\vec{\gamma},pp)$
asymmetries we display model predictions for a
((1p$_{3/2}$)$^{-2}$;$J_R=0^+,2^+$) and a 
((1p$_{3/2}$)$^{-1}$(1p$_{1/2}$)$^{-1}$;$J_R=1^+$) final state in 
Fig.~\ref{fig:monday}.  In comparison with the 
((1p$_{1/2}$)$^{-2}$;$J_R=0^+$) situation, the considered
shell-model configurations are not as selective as far as the different
contributing relative angular momentum states is concerned.
Indeed, the (1p$_{3/2}$)$^{-2}$ configuration can be
excited in proton-proton knockout through $S$, $P$ and $D$ absorption.
The ((1p$_{3/2}$)$^{-1}$(1p$_{1/2}$)$^{-1}$;$J_R=1^+$) configuration
is unique in that $S$-wave absorption is excluded and 
only relative $P$ waves would contribute in a
Harmonic Oscillator model \cite{gerco}.  From the right panel of
Fig.~\ref{fig:monday} it becomes clear that this produces a
doubly-bumped structure in the $\theta_p$ dependence with a minimum
around $P=0$.  At the same time, the asymmetry is positive in most of
the covered ($E_{\gamma},\theta_p$) region.  The left panel of
Fig.~\ref{fig:monday} shows features that turn out  to be a hybrid mixture
of characteristics related to $S$ and $P$ wave absorption.  The $P$
wave admixture makes the asymmetry to be substantially smaller than
what was obtained in the left panel of Fig.~\ref{fig:sunday} where the
$S$ wave dominates.  Rather strong variations of $\Sigma$ in E$_{\gamma}$ and
$\theta _p$ are observed and there is very little resemblance with a
flat structure wiggling slightly above the $\Sigma$=-1 plane as was
noticed for the $(\gamma,pp)$ results in Fig.~\ref{fig:sunday}.  So,
rather than FSI effects or Fermi motion it turns out that
photoabsorption on relative $P$ diprotons has a very large impact on
the ($\vec{\gamma}$,$pp$) asymmetries.
  The present results seem to suggest that an admixture of $P$-wave
absorption might also help in interpreting the measured 
$^{3}$He($\vec{\gamma},pp$)
asymmetries (Ref.~\cite{tede}) which were shown to be incompatible
with $^1 S_0$ photoabsorption in Ref.~\cite{sandorfi}. 
  

In an attempt to study the photonenergy dependence of the
$(\gamma,pp)$ variables more profoundly and estimate the effect of the
ground-state correlations, we display in Fig.~\ref{fig:coppsea} some
cuts along fixed values of the opening angle in the three-dimensional
plot of Fig.~\ref{fig:sunday}.  The opening angles are chosen to lie
in that part of the phase space where the cross sections are
reasonably large.  Four values of the opening angles have been
considered.  The sensitivity of the cross sections to the ground-state
correlations is found to be relatively small.  So is the effect on the
asymmetry for most of the phase space covered. However, the spike
observed at $\theta_p$=90$^o$ and lower photon energies (see also
Fig.~\ref{fig:sunday}) has to be fully ascribed to the ground-state
correlations.  It is maybe worth stressing that $\theta_p$=90$^o$
corresponds with the situation that the largest values of the relative
pair momentum are reached in coplanar and symmetrical kinematics.  The
effect of the short-range effects on the proton polarization is large.
This might however be very characteristic for the chosen shell-model
configuration in the final state.  Indeed, the proton polarization turns out
to be zero when considering only the isobaric contribution and a 0$^+$
final state.

The corresponding $(\gamma,pn)$ results for the proton-proton results
in Fig.~\ref{fig:coppsea} are contained in Fig.~\ref{fig:coppnsea}.
For all nucleon emission angles considered, a wide $\Delta$ resonant
structure is visible in the cross section.  Surprisingly, the resonant
behaviour is not pronounced for the polarization observables.
This confirms the conclusion that the
asymmetry is extremely sensitive to the interference terms between the
different contributing two-body operators.  Remark that the predicted
effect of central ground-state correlations on the $(\gamma,pn)$ observables
is marginal.

\section{Conclusions and prospects}
\label{sec:conclus}
In this paper we have explored the possibilities of using polarized
photon beams and outgoing nucleon polarimetry to learn more about the
dynamics of bound nucleon pairs with the aid of $(\gamma,pp)$ and
$(\gamma,pn)$ reactions.  As it is often the case, the asymmetry was
shown to be less dependent on the outgoing nucleon distortions than
the corresponding differential cross sections.  The effect of the outgoing
nucleon 
distortions on the polarization, however, is rather large.  The
$P_y^p$ was further shown to be a sensitive observable to
investigate the mechanisms related to the short lifetime of the
$\Delta$ resonance in the medium.  Indeed, the latter
reflect themselves in imaginary parts entering the $\Delta$ selfenergy
and produce sizeable contributions to the outgoing nucleon polarization.

It was pointed out that by studying the differential cross sections
for excitation of the $A-2$ fragment in a state with particular
angular momentum $J_R$ one can deduce information about the nature of
the initial pair wave function.  This does not imply, however, that
the experiments would have to resolve each residual state.
Observables for different shell combinations, which would typically
feed the residual system in a certain range of missing energies, would
already be very instructive to learn about the different possible pair
combinations.  We have shown that the observables are characterized by
a selective sensitivity to mechanisms that go beyound photoabsorption
on a $^1S_0$ proton-proton and $^3S_1$ proton-neutron pair.  This was
found to be already the case in so-called quasi-deuteron kinematics
where the missing momentum is constrained to be zero, so that the
influence of the pair c.m. motion is minimized and ``deuteron-like''
conditions are created.  In this type of kinematics, the major
deviations from S-wave absorption are visible in the angular cross
sections.  The polarization observables seem to be far less affected.
The fact that S-wave absorption plays a predominant role does not
guarantee, however, that the $(\gamma,pn)$ asymmetries exhibit
deuteron-like properties in quasi-deuteron kinematics.  In the
$\Delta$ resonance region a strong similarity between the
d$(\vec{\gamma},pn)$ and the $^{12}$C$(\vec{\gamma},pn)$ asymmetries
is observed.  The situation changes for the lower photon energies
where the $^{12}$C predictions deviate substantially from the measured
deuteron asymmetries.  For all these reasons, we deem that the
exclusive $(\gamma,NN)$ channel is an ideal probe to study the
limitations of the QD approach and to reach a better level in our
understanding of the dynamics of proton-neutron pairs in the medium
relative to the deuteron.


Within the factorized model, a general behaviour for the
A$(\vec{\gamma},pp)$ asymmetries is predicted in that they would be
exactly -1 as long as one is dealing with $^1S_0$ absorption in
coplanar kinematics and the dominant role of the isobar current is
guaranteed.  Neither FSI effects, nor a more sophiscated treatment of
the pair relative $S$ wave function seem to change this very much.
The slightest admixture of mechanisms going beyond $^1S_0$ absorption
are noticed to induce major changes to the asymmetry.  Whereas $\Sigma
\approx$=-1 for $^1 S _0$ photoabsorption, the calculations
predict a completely different behaviour as soon as relative $P,D,...$ start
playing a role. The extent to which mechanisms beyound $S$ wave
absorption play a role is very much dependent on the
kinematics and the nuclear structure of the A-2 fragment.  The
sensitivity of the ($\gamma,pp$) asymmetries to ground-state
correlations was shown not to be very large and was considerably
smaller than the effects on the angular cross sections and outgoing
nucleon polarizations.  Such a behaviour could already be inferred
from the expressions derived within the context of the factorized
model.  All this, however, is illustrating the usefulness of having
simplified models for predicting the major trends and sensitivities in
the different observables.

It would be of interest to see to which degree the deviations from
QD-like absorption and deuteron-like behaviour can be 
experimentally confirmed.  In any case, high or moderate resolution
data in the low-energy part of the missing energy spectrum would
provide invaluable information to test wether the dynamics of the
pairs in the medium complies with mean-field like behaviour or whether
there are important deviations in (some of) the relative wave function
combinations.  In order to minimize the uncertainties regarding the
final state interaction polarization observables, and in particular
the photon asymmetries, will be of very great help in these studies.

\section*{Acknowledgment}
This work was supported by the Fund for Scientific Research -
Flanders (FWO) and in part by the NATO through the
research grant CRG970268. One of us (J.R.) thanks the Institute for
Nuclear Theory at the University of Washington for partial support
during the completion of this work.

\newpage
\begin{table}
\caption{Possible configurations for proton-proton knockout from the
  s- and p-shell combinations. 
  The $\Lambda$ (l) denotes the c.m. (relative) angular momentum of
  the pair. The J$_{pair}^{\pi}$ is the total angular momentum of the
  pair (including relative and c.m. motion). The separation in
c.m. and relative motion is done as if we are dealing with HO
single-particle wave functions.}
\label{tab:pp} 
\begin{tabular}{l|c|c|c|c}\hline
shell-model &  & relative & c.m.  &
relative wave function  \\
(n$_h$,l$_h$)(n$_{h'}$,l$_{h'}$) & (L,S)J$_{pair}^{\pi}$  & (n,l)  
& (N,$\Lambda$) & $^{2S+1}l_{J}$ (T)    \\
\hline
(1s)$^2$         & $(0,0)0^+$ & (1,0) & (1,0) & $^1S_0$ (T=1) \\ \hline
(1s)(1p)         & $(1,1)0^-$ & (1,1) & (1,0) & $^3P_0$ (T=1) \\
                 & $(1,0)1^-$ & (1,0) & (1,1) & $^1S_0$ (T=1) \\
                 & $(1,1)1^-$ & (1,1) & (1,0) & $^3P_1$ (T=1) \\
                 & $(1,1)2^-$ & (1,1) & (1,0) & $^3P_2$ (T=1) \\ \hline
(1p)$^2$         & $(0,0)0^+$ & (1,0) & (2,0) & $^1S_0$ (T=1) \\ 
                 & $(0,0)0^+$ & (2,0) & (1,0) & $^1S_0$ (T=1) \\
                 & $(1,1)0^+$ & (1,1) & (1,1) & $^3P_1$ (T=1) \\
                 & $(1,1)1^+$ & (1,1) & (1,1) & $^3P_0,^3P_1,^3P_2$ (T=1) \\
                 & $(1,1)2^+$ & (1,1) & (1,1) & $^3P_1,^3P_2$ (T=1) \\
                 & $(2,0)2^+$ & (1,0) & (1,2) & $^1S_0$ (T=1) \\
                 & $(2,0)2^+$ & (1,2) & (1,0) & $^1D_2$ (T=1) \\ \hline
\end{tabular}
\end{table}

\begin{table}
\caption{Possible configurations for proton-neutron knockout from the
  s- and p-shell combinations. The $\Lambda$ (l) denotes the
  c.m. (relative) angular momentum of
  the pair. The J$_{pair}^{\pi}$ is the total angular momentum of the
  pair (including relative and c.m. motion).  The separation in
c.m. and relative motion is done as if we are dealing with HO
single-particle wave functions.}
\label{tab:pn} 
\begin{tabular}{l|c|c|c|c}\hline
shell-model &  & relative & c.m.  &
relative wave function  \\
(n$_h$,l$_h$)(n$_{h'}$,l$_{h'}$) & (L,S)J$_{pair}^{\pi}$  & (n,l)  
& (N,$\Lambda$) & $^{2S+1}l_{J}$ (T)    \\
\hline
(1s)$^2$         & $(0,0)0^+$ & (1,0) & (1,0) & $^1S_0$ (T=1)\\ 
                 & $(0,1)1^+$ & (1,0) & (1,0) & $^3S_1$ (T=0)\\ \hline
(1s)(1p)         & $(1,1)0^-$ & (1,0) & (1,1) & $^3S_1$ (T=0)\\
                 & $(1,1)0^-$ & (1,1) & (1,0) & $^3P_0$ (T=1)\\
                 & $(1,0)1^-$ & (1,1) & (1,0) & $^1P_1$ (T=0)\\
                 & $(1,0)1^-$ & (1,0) & (1,1) & $^1S_0$ (T=1)\\
                 & $(1,1)1^-$ & (1,0) & (1,1) & $^3S_1$ (T=0)\\
                 & $(1,1)1^-$ & (1,1) & (1,0) & $^3P_1$ (T=1)\\
                 & $(1,1)2^-$ & (1,0) & (1,1) & $^3S_1$ (T=0)\\ 
                 & $(1,1)2^-$ & (1,1) & (1,0) & $^3P_2$ (T=1)\\ \hline
(1p)$^2$         & $(0,0)0^+$ & (2,0) & (1,0) & $^1S_0$ (T=1)\\ 
                 & $(0,0)0^+$ & (1,0) & (2,0) & $^1S_0$ (T=1)\\
                 & $(1,1)0^+$ & (1,1) & (1,1) & $^3P_1$ (T=1)\\
                 & $(0,1)1^+$ & (2,0) & (1,0) & $^3S_1$ (T=0)\\ 
                 & $(0,1)1^+$ & (1,0) & (2,0) & $^3S_1$ (T=0)\\
                 & $(1,1)1^+$ & (1,1) & (1,1) & $^3P_0,^3P_1,^3P_2$ (T=1)\\                  
                 & $(1,0)1^+$ & (1,1) & (1,1) & $^1P_1$ (T=0)\\ 
                 & $(2,1)1^+$ & (1,0) & (1,2) & $^3S_1$ (T=0)\\
                 & $(2,1)1^+$ & (1,2) & (1,0) & $^3D_1$ (T=0)\\
                 & $(1,1)2^+$ & (1,1) & (1,1) & $^3P_1,^3P_2$ (T=1)\\
                 & $(2,0)2^+$ & (1,0) & (1,2) & $^1S_0$ (T=1)\\
                 & $(2,0)2^+$ & (1,2) & (1,0) & $^1D_2$ (T=1)\\
                 & $(2,1)2^+$ & (1,0) & (1,2) & $^3S_1$ (T=0)\\ 
                 & $(2,1)2^+$ & (1,2) & (1,0) & $^3D_2$ (T=0)\\
                 & $(2,1)3^+$ & (1,0) & (1,2) & $^3S_1$ (T=0)\\
                 & $(2,1)3^+$ & (1,2) & (1,0) & $^3D_3$ (T=0)\\ 
\end{tabular}
\end{table}

\begin{figure} [h]
\caption{Basic diagrams contributing to photopion production on the
nucleon. The diagrams are drawn in a rather uncommon fashion so as to
make the link with photoinduced $2N$ knockout more clear (Fig.~
\protect \ref{fig:twon}). (a) Kroll-Rudermann term ; (b) pion-pole
term ; (c,f) direct and crossed nucleon Born terms ; (d,e) direct and
crossed $\Delta$ terms.}
\label{fig:pion}
\end{figure}

\begin{figure} [h]
\caption{The equivalent diagrams of Fig.~\protect\ref{fig:pion} for
two-nucleon knockout from finite nuclei in the spectator approximation. 
Each solid line shows a nucleon moving in a mean-field potential.}
\label{fig:twon}
\end{figure}

\begin{figure}[ht]
\caption{Unpolarized differential cross section, photon asymmetry and 
proton polarization for
the $^{12}$C($\gamma,pp$)$^{10}$Be((1p$_{3/2}$)$^{-2}$) at
E$_{\gamma}$=100, 200, 300 and 400~MeV in QD
kinematics. The separate contribution from $J_R$=0$^+$ (dashed line)
and $J_R$=2$^+$ (dotted line) are shown.  The solid line is the
incoherent sum of both contributions. The curves are the result of
an unfactorized calculation including outgoing nucleon distortions,
isobaric currents and ground-state correlations.  The latter are
implemented with the central correlation function from Ref. \protect \cite{gd}.}
\label{fig:seapp}
\end{figure}

\begin{figure}[ht]
\caption{Unpolarized differential cross section, photon asymmetry 
and proton polarization for
the $^{12}$C($\gamma,pn$)$^{10}$B((1p$_{3/2}$)$^{-2}$) reaction at
E$_{\gamma}$=100, 200, 300 and 400~MeV in QD
kinematics. The separate contribution from $J_R$=0$^+$ (dashed line),
$J_R$=1$^+$ (dot-dashed line) and $J_R$=3$^+$ (dotted line) are shown.
The $J_R$=2$^+$ component is small and has been omitted from the
figure. 
The solid line is the incoherent sum of all $J_R$ components.
The curves are the result of
an unfactorized calculation including outgoing nucleon distortions,
pion exchange currents, 
isobaric currents and ground-state correlations.  The latter are
implemented with the central correlation function from Ref. \protect \cite{gd}.
The asymmetries are compared with d($\vec{\gamma},p)n$ data from
Refs.~\protect \cite{legsdata} (triangles) and 
\protect \cite{russian} (squares). The proton polarization is compared
with d($\gamma,\vec{p})$n data from Ref.~\protect \cite{ganenko}.}
\label{fig:seapn}
\end{figure}

\begin{figure}
\caption{Unpolarized differential cross section, photon asymmetry and 
proton polarization for the $^{16}$O($\gamma,pn$) reaction for two
types of p-shell configurations in QD kinematics and 
E$_{\gamma}$=150~MeV. The dashed line is obtained when including
only the seagull pion-exchange current. The dotted line is the
calculated result when including both the seagull and
pion-in-flight pion-exchange current.  The solid line shows
the full model calculation, including the two pion exchange and the $\Delta
_{33}$ currents. The dot-dashed line shows the equivalent result 
but now using plane
waves for the outgoing particle wave functions.}
\label{fig:figa}
\end{figure}

\begin{figure}
\caption{Differential cross section, asymmetry and polarization for
the $^{12}$C($\gamma,pp$)$^{10}$Be((1p$_{3/2}$)$^{-2}$) reaction at
E$_{\gamma}$=100 and 300~MeV in QD
kinematics. The dotted curves are the predictions of the factorized model.
The solid (dashed) line are the results of the unfactorized model
using distorted  (plane) outgoing nucleon waves.   
In all cases
isobaric currents and ground-state correlations are included.  The latter are
implemented with the central correlation function from Ref. \protect \cite{gd}.}
\label{fig:overpp}
\end{figure}

\begin{figure}
\caption{The $^{12}$C($\gamma,pp$) observables in QD kinematics for
  knockout from the p-shell. All curves are obtained in the
  unfactorized model and include outgoing nucleon distortions. The
  dashed line includes solely the isobaric currents.  The other curves
  include also Jastrow correlations with different choices for the
  correlation function : Ref.~\protect \cite{gd} (solid),
  Ref.~\protect \cite{euler} (dotted) and Ref.~\protect \cite{omy} 
(dot-dashed).}
\label{fig:wasrc}
\end{figure}

\begin{figure}
\caption{Photon energy and proton emission angle dependency of the 
  $^{16}$O($\gamma,pp$)(1p$_{1/2}$)$^{-2}$ and
  $^{16}$O($\gamma,pn$)(1p$_{1/2}$)$^{-2}$ differential cross section and
  photon asymmetry in coplanar and symmetrical kinematics. The calculations
  include the outgoing nucleon distortions and all contributing
  meson-exchange and isobaric current diagrams.  Ground-state
  correlations were implemented through the correlation function from
  Ref.~\protect \cite{gd}.}
\label{fig:sunday}
\end{figure}

\begin{figure}
\caption{Photon energy and emission angle dependency of the 
$^{16}$O($\gamma,pp$)$^{14}$C differential cross section
and asymmetry in coplanar and symmetrical kinematics. Two different
shell-model configurations are considered.  The calculations
include the outgoing nucleon distortions and all contributing
isobaric current diagrams.  
Also the ground-state
correlations were implemented using the correlation function from
Ref.~\protect \cite{gd}.} 
\label{fig:monday}
\end{figure}

\begin{figure}[ht]
\caption{Photon energy dependency of cross section,
asymmetry and polarization for
the $^{16}$O($\gamma,pp$)$^{14}$C((1p$_{1/2}$)$^{-2}$,$J_R=0^+$) reaction 
in coplanar
and symmetrical kinematics and different values for the opening angle 
$\theta _p$. 
The curves are the result of
an unfactorized calculation including outgoing nucleon distortions.  The dashed
curves are the results when including solely isobaric currents.  For the solid 
lines also the effect of ground-state correlations are included.  The latter are
implemented with the central correlation function from Ref. \protect \cite{gd}.}
\label{fig:coppsea}
\end{figure}

\begin{figure}[ht]
\caption{Photon energy dependency of cross section,
asymmetry and polarization for the
$^{16}$O($\gamma,pn$)$^{14}$N((1p$_{1/2}$)$^{-2}$)
reaction in coplanar and symmetrical kinematics and different values
for the opening angle $\theta _p$=$\theta _n$. The 
contributions from $J_R=0^+$ and $1^+$ are incoherently added.  
The dashed curves are the result of an
unfactorized calculation including outgoing nucleon distortions,
meson exchange and isobaric currents.  For the solid lines also the effect of
ground-state correlations are included.  The latter are implemented
with the central correlation function from Ref. \protect \cite{gd}.}
\label{fig:coppnsea}
\end{figure}


\newpage
\begin{figure}
\begin{center}
{\mbox {\epsfxsize=16.cm \epsffile{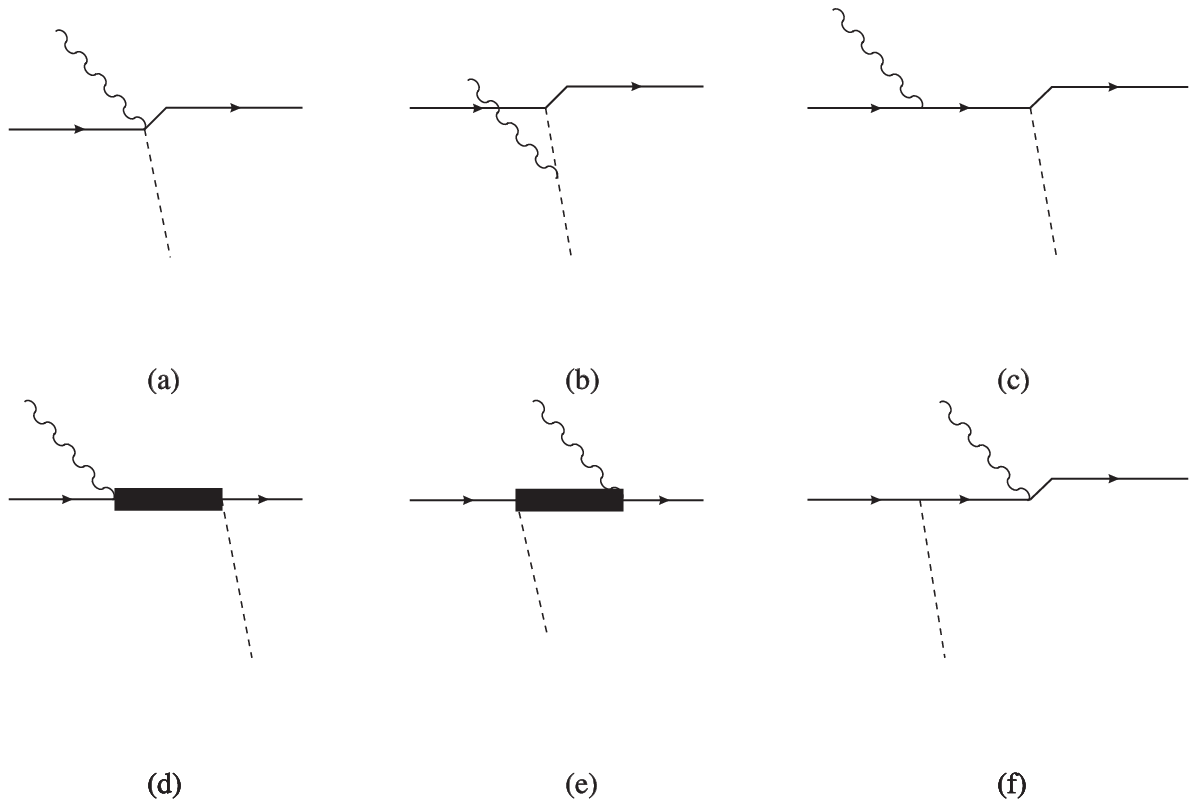}}}
\end{center}
\end{figure}
{\Huge Figure 1}

\newpage
\begin{figure}
\begin{center}
{\mbox {\epsfxsize=16.cm \epsffile{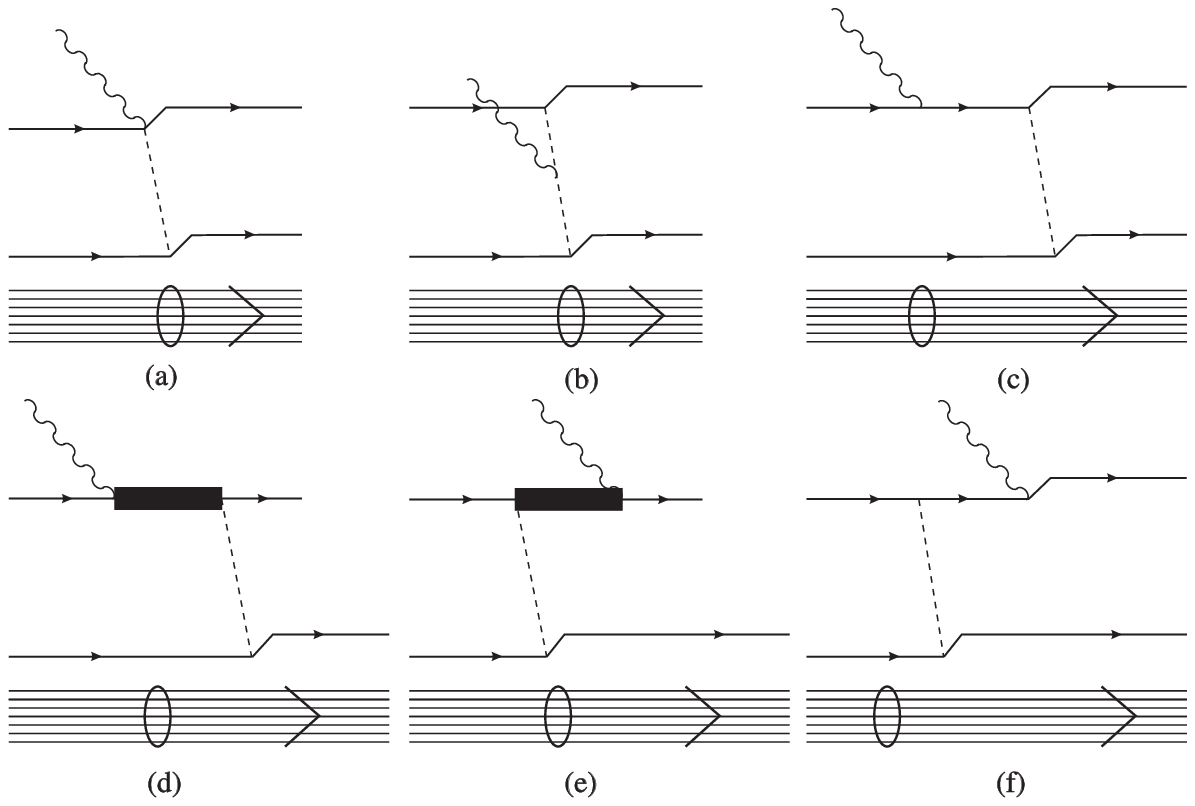}}}
\end{center}
\end{figure}
{\Huge Figure 2}

\newpage
\begin{figure}
\begin{center}
{\mbox {\epsfxsize=16.cm \epsffile{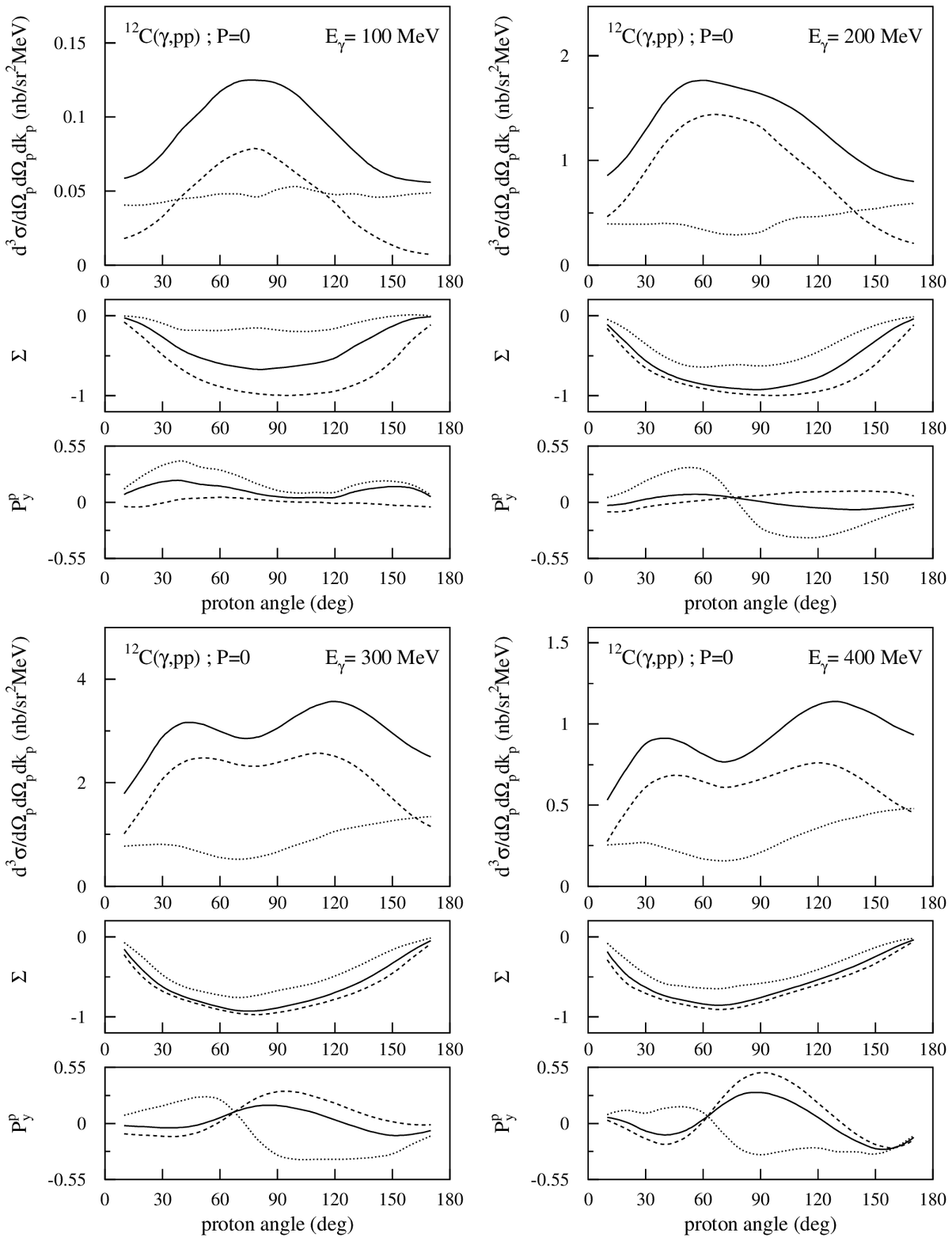}}}
\end{center}
\end{figure}
{\Huge Figure 3}

\newpage
\begin{figure}
\begin{center}
{\mbox {\epsfxsize=16.cm \epsffile{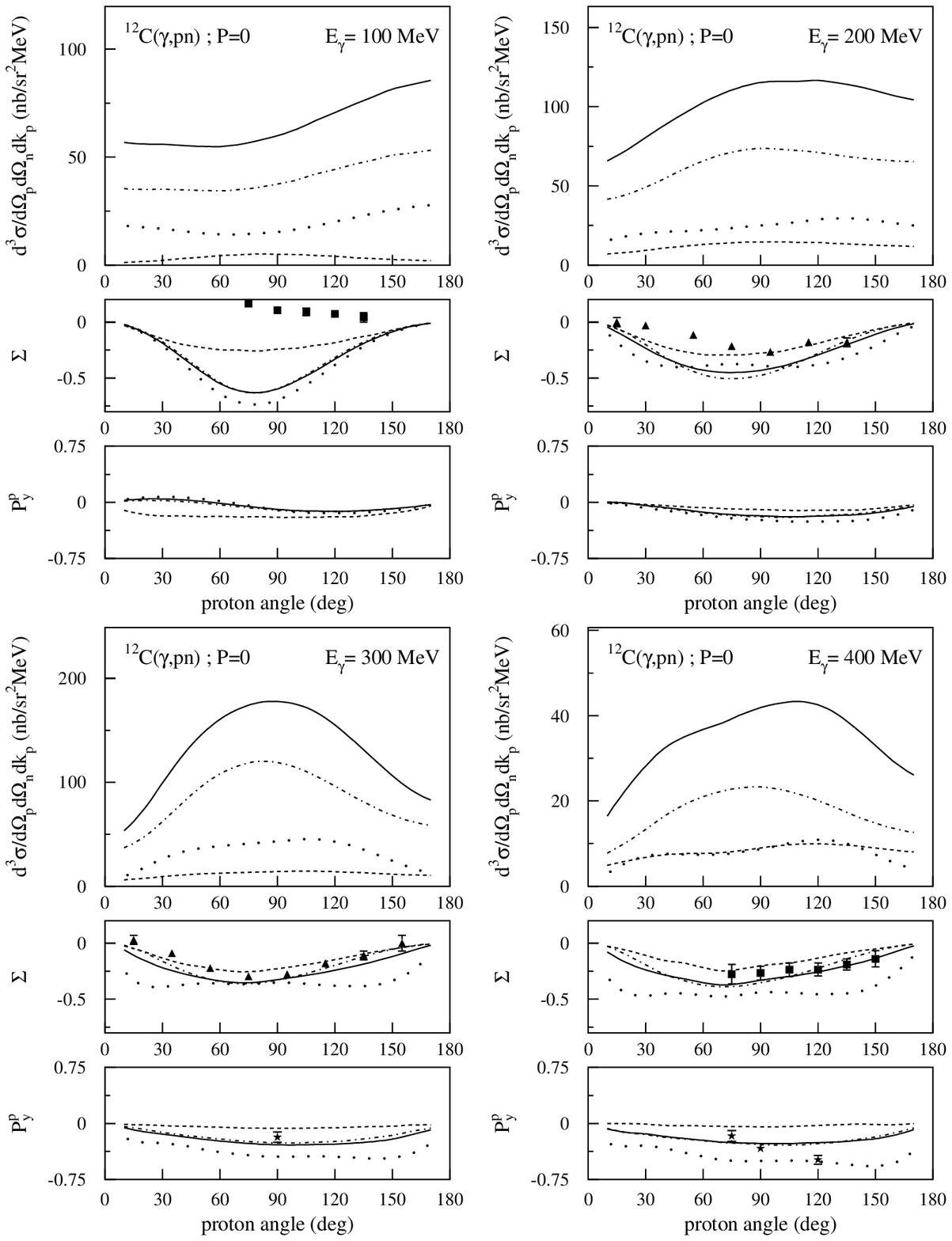}}}
\end{center}
\end{figure}
{\Huge Figure 4}

\newpage
\begin{figure}
\begin{center}
{\mbox {\epsfxsize=16.cm \epsffile{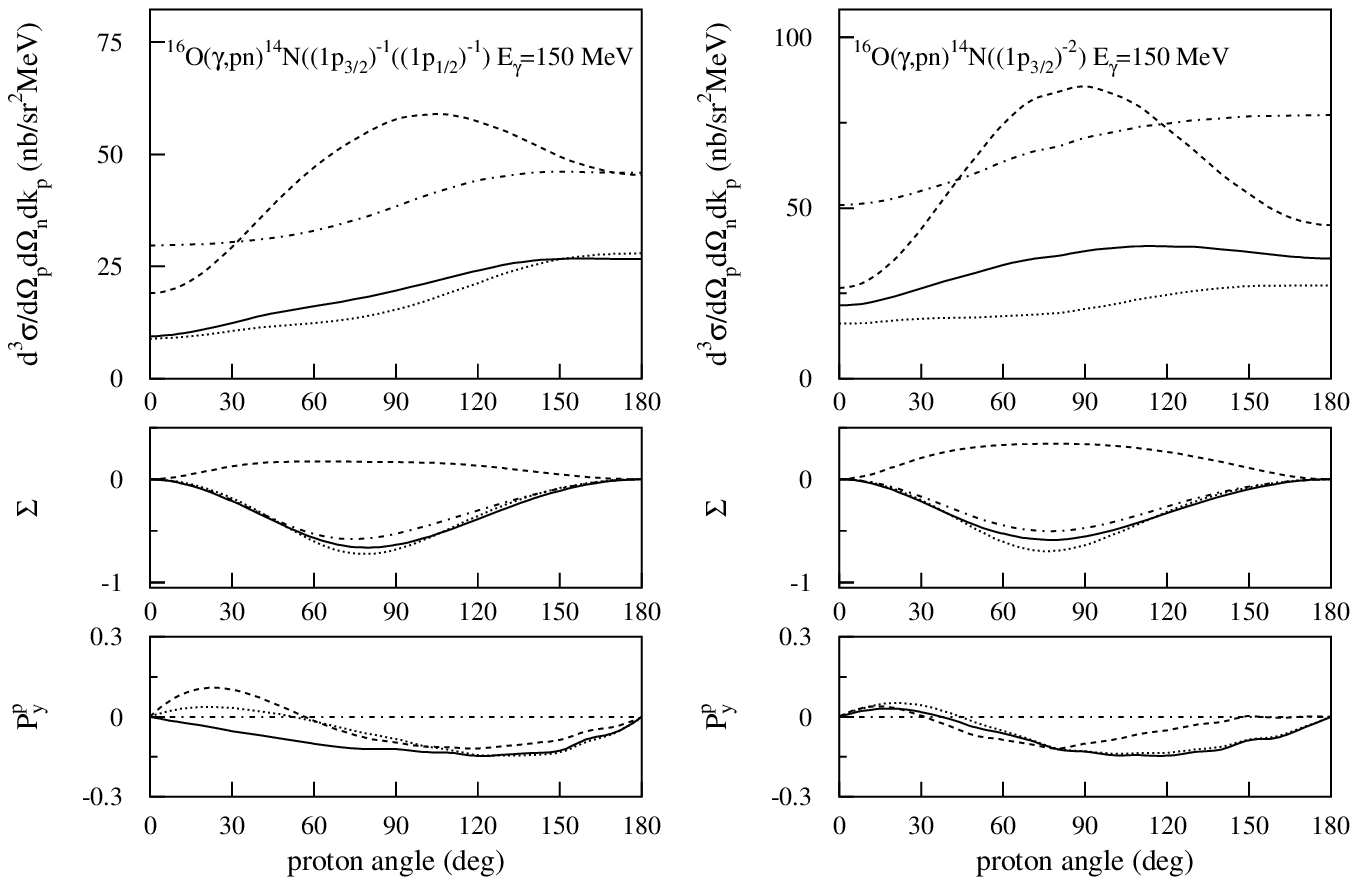}}}
\end{center}
\end{figure}
{\Huge Figure 5}

\newpage
\begin{figure}
\begin{center}
{\mbox {\epsfxsize=16.cm \epsffile{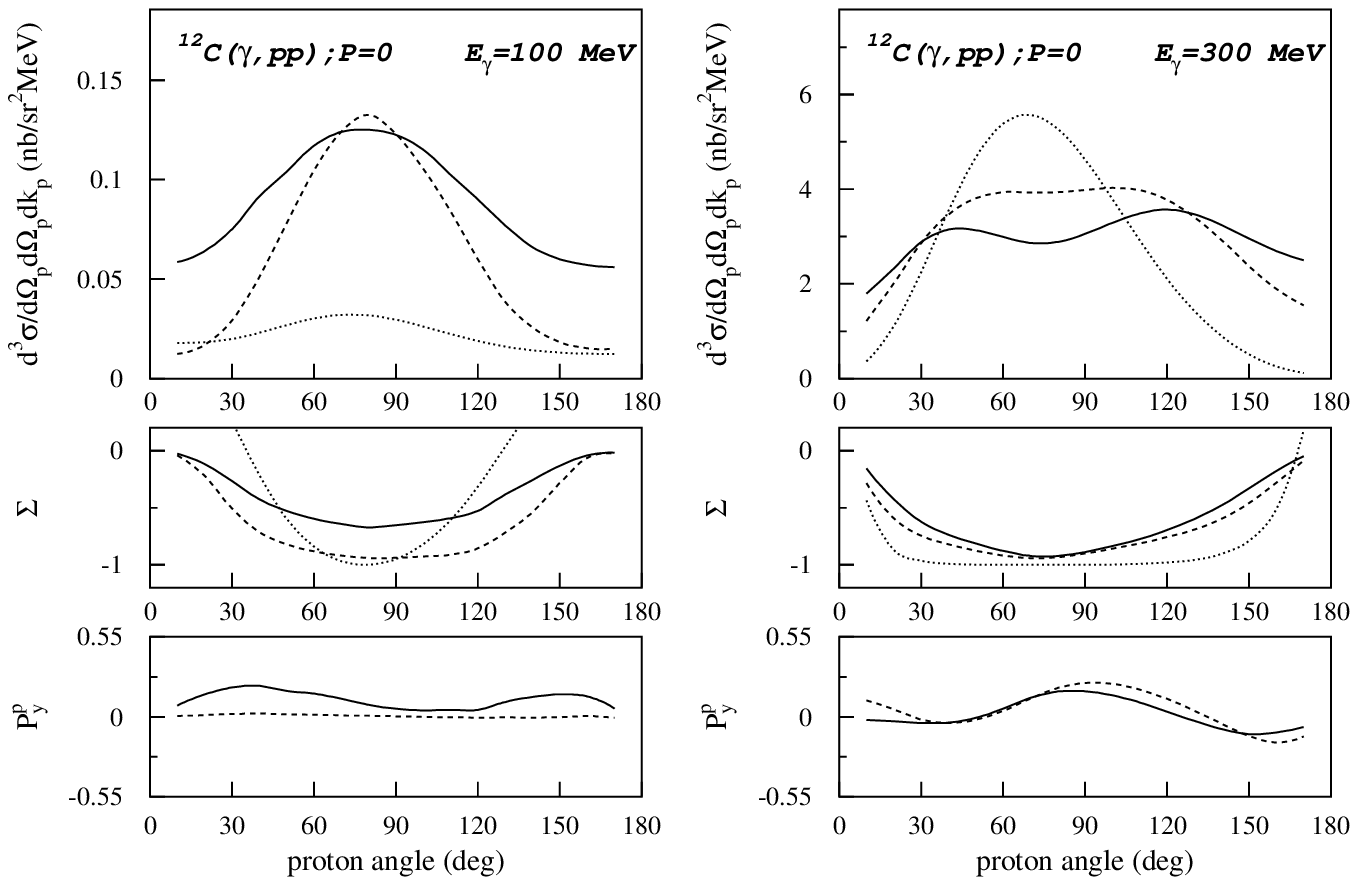}}}
\end{center}
\end{figure}
{\Huge Figure 6}

\newpage
\begin{figure}
\begin{center}
{\mbox {\epsfxsize=16.cm \epsffile{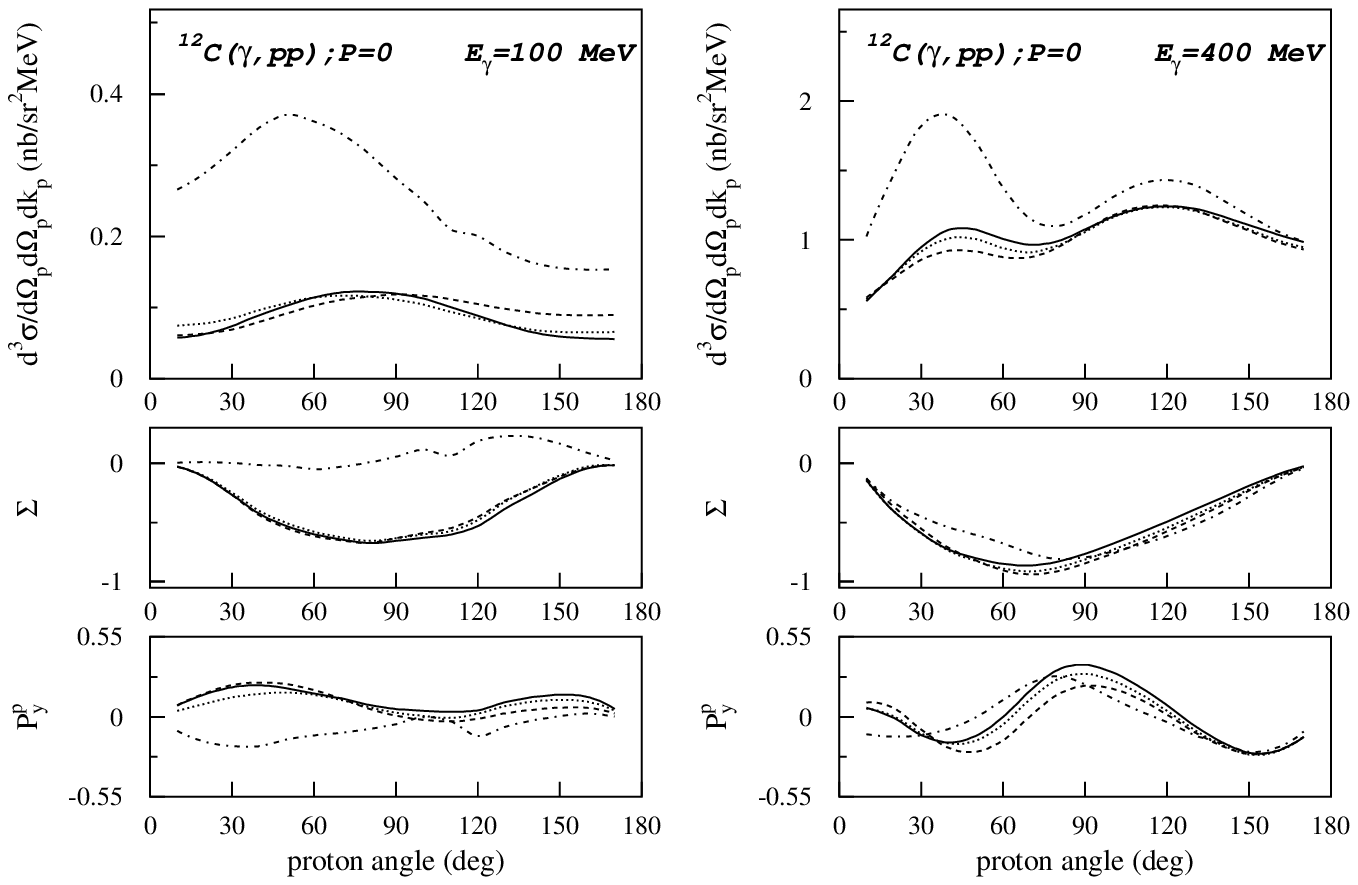}}}
\end{center}
\end{figure}
{\Huge Figure 7}

\newpage
\begin{figure}
\begin{center}
{\mbox {\epsfxsize=16.cm \epsffile{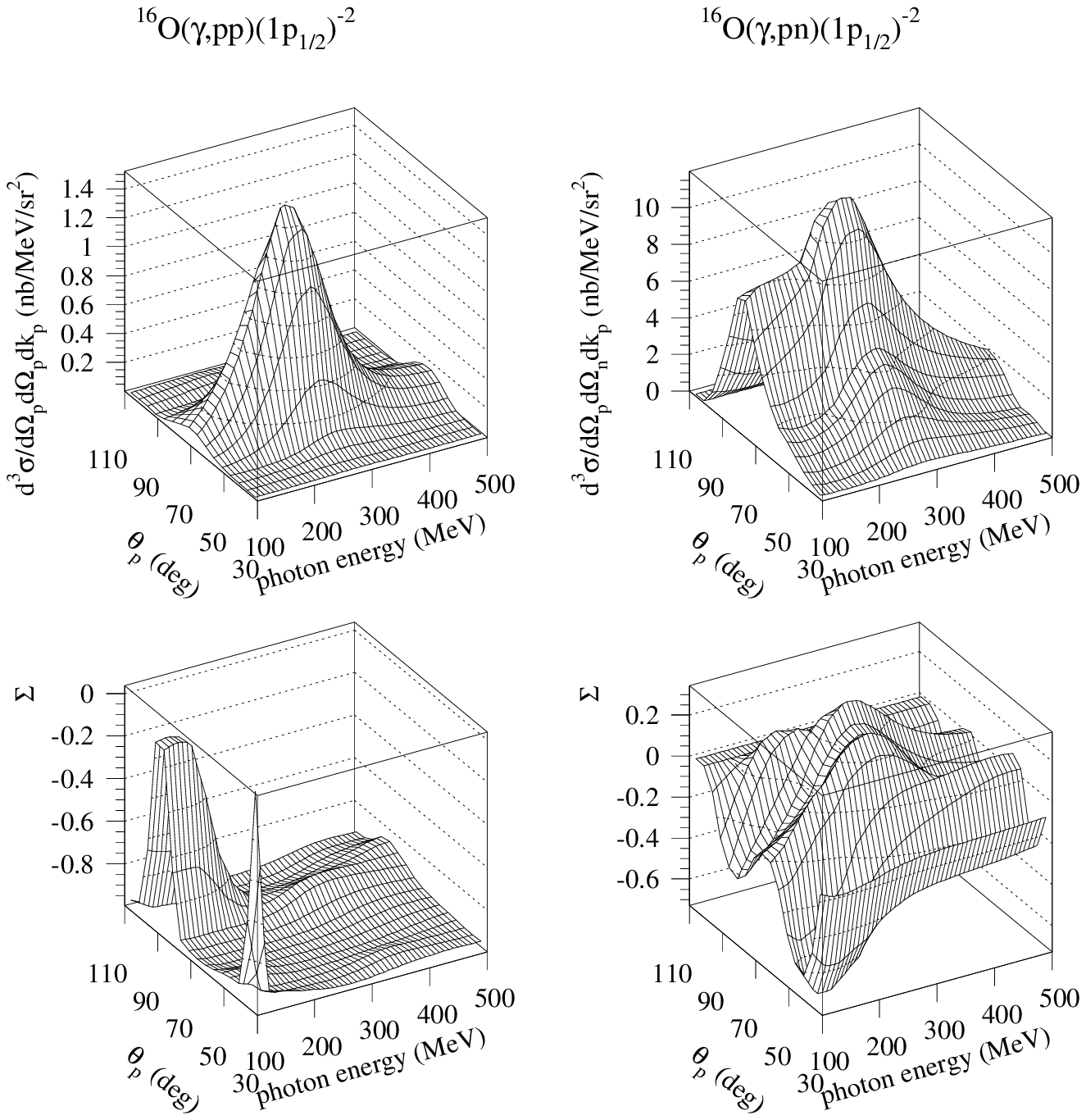}}}
\end{center}
\end{figure}
{\Huge Figure 8}

\newpage
\begin{figure}
\begin{center}
{\mbox {\epsfxsize=16.cm \epsffile{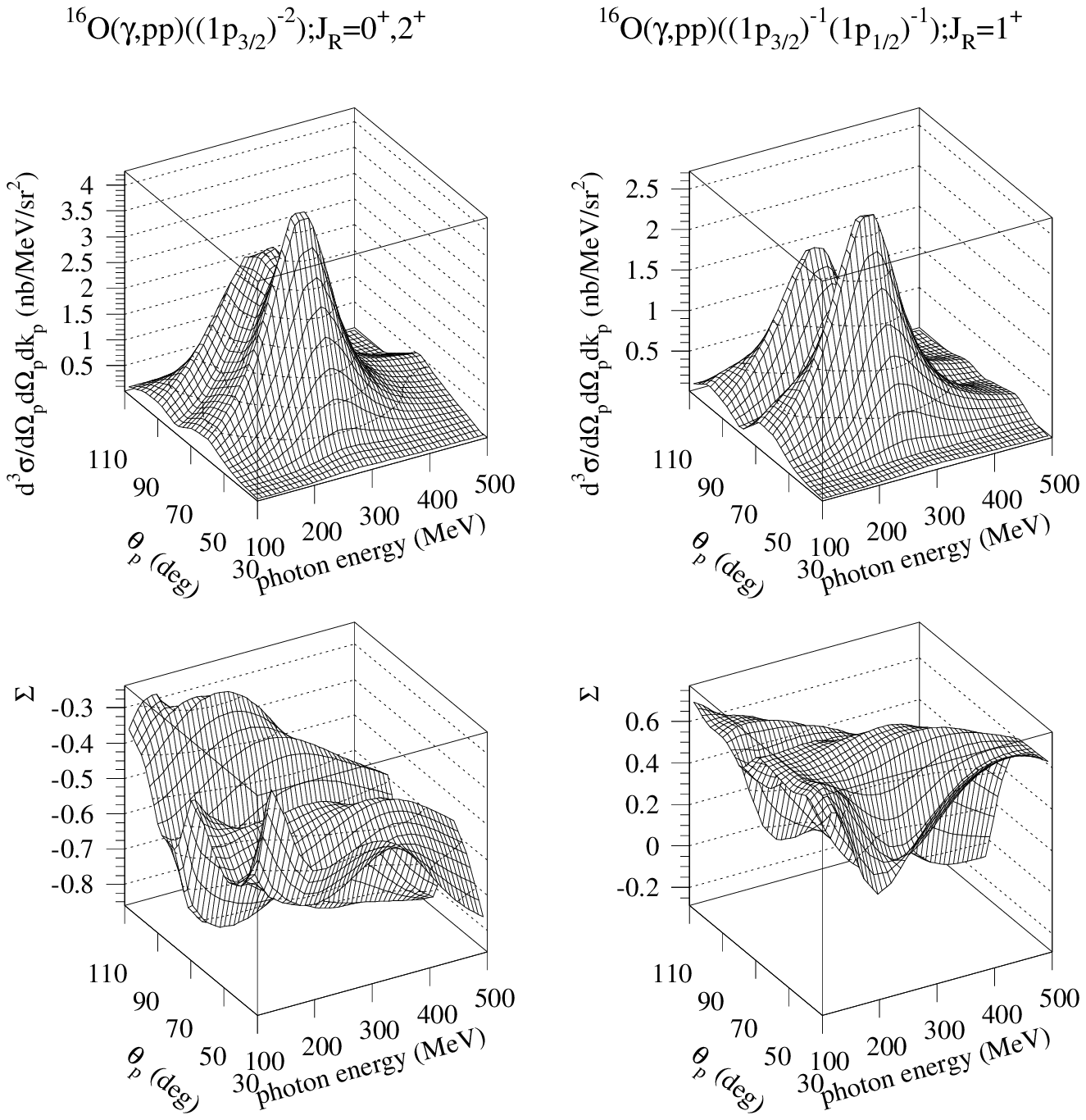}}}
\end{center}
\end{figure}
{\Huge Figure 9}

\newpage
\begin{figure}
\begin{center}
{\mbox {\epsfxsize=16.cm \epsffile{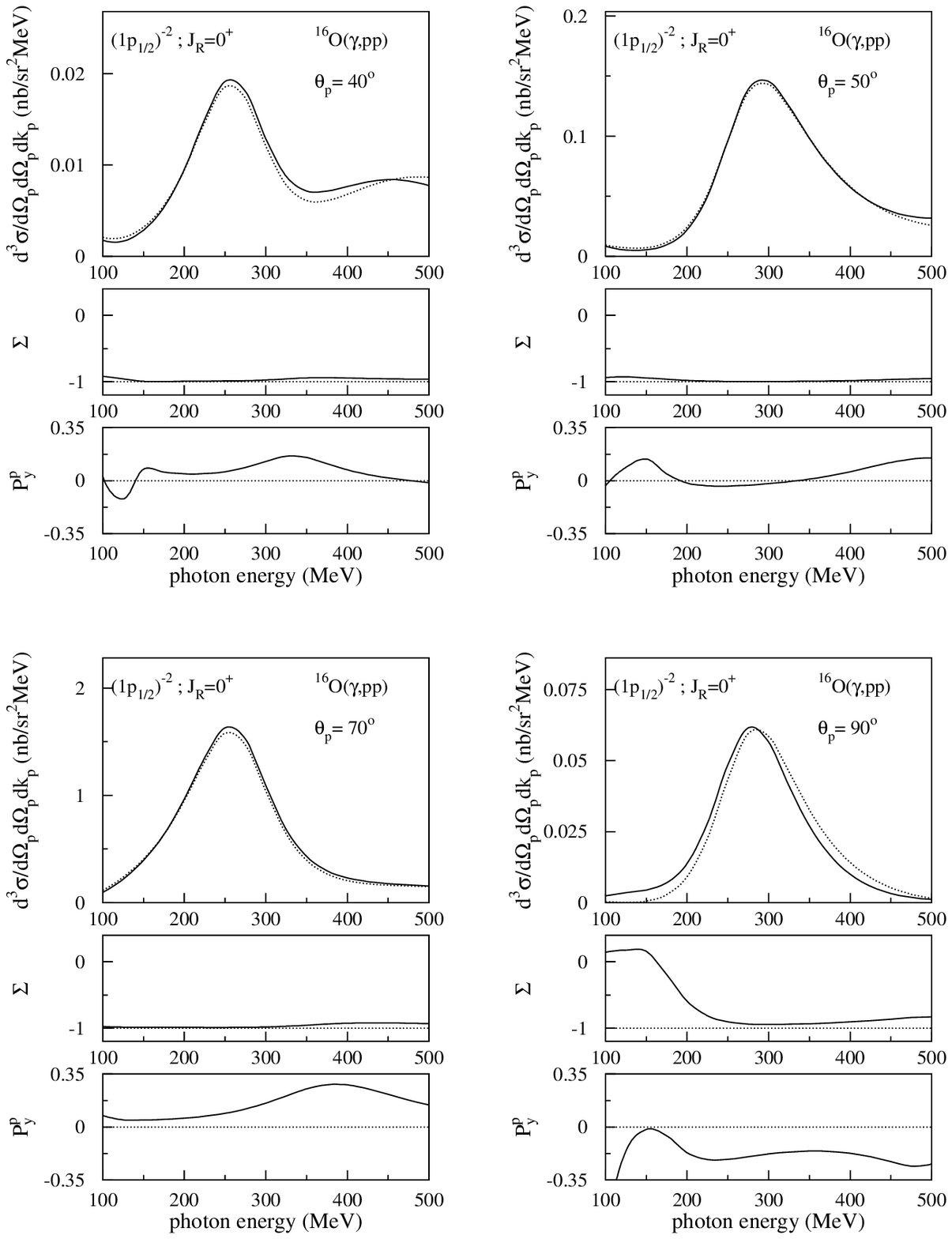}}}
\end{center}
\end{figure}
{\Huge Figure 10}

\newpage
\begin{figure}
\begin{center}
{\mbox {\epsfxsize=16.cm \epsffile{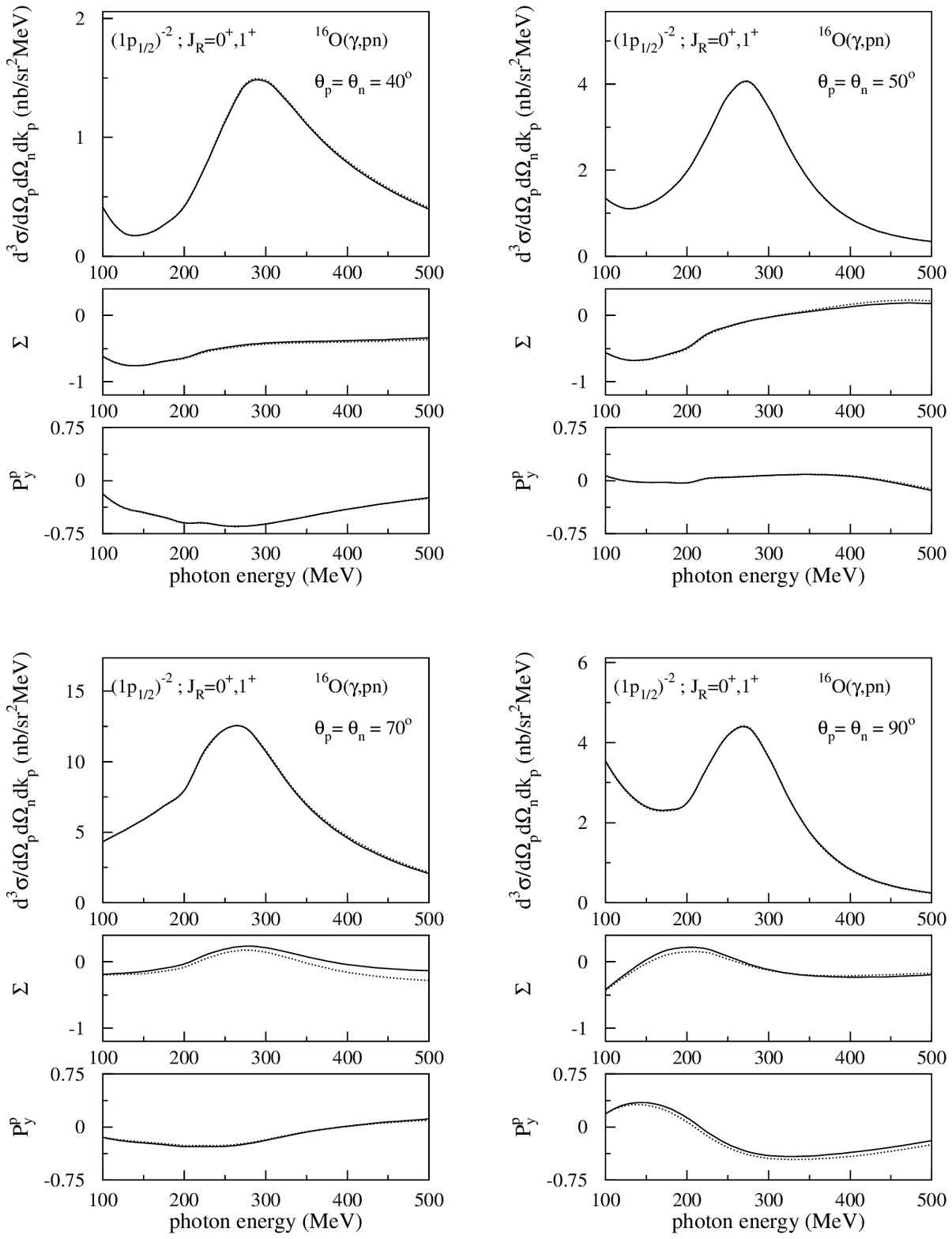}}}
\end{center}
\end{figure}
{\Huge Figure 11}

\end{document}